\documentclass[letterpaper]{article}

\usepackage[T1]{fontenc}

\usepackage{geometry}
\usepackage{setspace}
\usepackage[style = chem-acs]{biblatex}
\usepackage[dvipsnames]{xcolor} 
\usepackage{graphicx} 
\usepackage{subcaption} 
\usepackage{float}
\usepackage{amsmath, amssymb} 
\usepackage{bm} 
\usepackage{xr} 
\usepackage[normalem]{ulem} 

\usepackage{placeins}

\usepackage{listings}
\lstdefinelanguage{InputScript}{
  morekeywords={System, Selection, Analysis, Create, Type, File, TimeScheme, Output},
  sensitive=false,
  morecomment=[l]{\#},
  morestring=[b]",
}
\usepackage{xurl}
\usepackage{hyperref}
\hypersetup{
      colorlinks=true,
      linkcolor=blue,
      filecolor=magenta,      
      urlcolor=blue,
      citecolor=black,
      pdftitle={AMDAT Paper Draft},
}

\usepackage{booktabs}
\usepackage{longtable,makecell}
\usepackage{rotating}
\usepackage{tabularx}

\usepackage{pdflscape}

\usepackage{enumitem}
\setlist[itemize]{noitemsep}

\usepackage{authblk}

\title{\textsf{AMDAT}: An Open-Source Molecular Dynamics Analysis Toolkit
for Supercooled Liquids, Glass-Forming Materials, and Complex Fluids}

\author[1]{Pierre Kawak}
\author[1,2]{William F. Drayer}
\author[3]{Bao Khanh Ma}
\author[1]{David S. Simmons*}

\affil[1]{Department of Chemical, Biological, and Materials Engineering,
University of South Florida, Tampa, Florida}

\affil[2]{Department of Materials Science and Engineering,
University of Pennsylvania, Philadelphia, Pennsylvania}

\affil[3]{Bellini College of Artificial Intelligence, Cybersecurity and Computing,
University of South Florida, Tampa, Florida}

\date{*Email: dssimmons@usf.edu}

\begin{document}

\maketitle

\begin{abstract}
AMDAT (Amorphous Molecular Dynamics Analysis Toolkit) is an open-source
C++ toolkit for post-processing molecular dynamics trajectories, focused
on high-performance static and dynamic analyses of amorphous, glassy,
and polymer materials, including supercooled liquids and complex fluids.
In this paper, we describe AMDAT's design for efficient long-timescale
analysis via in-memory trajectory handling and exponential time sampling,
and we demonstrate representative workflows for widely used observables
such as radial distribution functions, structure factors, intermediate
scattering functions, and neighbor correlations.
\end{abstract}

\section{Introduction}

The last 30 years have seen the maturation of molecular dynamics (MD) simulation methods, with the development of a substantial number of well-validated, open-source software packages for systems ranging from biomolecules to simple physics models~\cite{LAMMPS, GROMACS, NAMD, AMBER, CHARMM, OpenMM}.
These tools lower the barrier to entry for new practitioners, reduce duplicated effort by providing shared, extensible codebases, and enable published simulation work to be reproduced when authors report the precise software version and input configurations used.

In contrast, software for analysis of MD simulations has generally lagged behind simulation engines.
For much of the history of the field, many research groups have relied on heterogeneous collections of some analysis tools-some built into the simulation packages themselves, some integrated into visualization software, and some maintained as in-house codes.
As a result, published analyses are often difficult to reproduce exactly, many reported quantities rely on code that has not been broadly validated, and early-career researchers routinely spend substantial time re-implementing basic analysis functionality for their groups.

In recent years, several packages for MD analysis have been released.\cite{Hinsen_MMTK_2000,michaud-agrawal_mdanalysis_2011,gowers_mdanalysis_2016,seeber_wordom_2007, seeber_wordom_2011, felline_wordom_2023,romo_loos_2009, romo_lightweight_2014,roe_ptraj_2013, roe_parallelization_2018,mcgibbon_mdtraj_2015}
The most progress has been made in packages targeting analysis of protein and biomolecular simulations\cite{roe_ptraj_2013, roe_parallelization_2018,romo_loos_2009, romo_lightweight_2014} and in tools that provide a versatile coding environment for simulation analysis~\cite{VMD, OVITO, mcgibbon_mdtraj_2015,michaud-agrawal_mdanalysis_2011,gowers_mdanalysis_2016} (e.g., python-based tools for trajectory file reading and atom selection) rather than in broadly validated implementations of complete analysis methods themselves.
This presents a particular limitation for molecular simulations in the fields of soft matter physics, polymer physics, and supercooled liquids, and in studies of glass formation, where properties depend on subtle structural and dynamical correlations.\cite{cavagna_supercooled_2009}
For these systems, there remains a paucity of widely adopted, well-tested software for calculating the structural and dynamic properties most relevant to polymer structure and dynamics, glass formation, and spatially heterogeneous properties.

In this paper, we introduce the Amorphous Molecular Dynamics Analysis Toolkit (\textsf{AMDAT}), an open-source molecular dynamics trajectory analysis toolkit designed to fill this gap.
\textsf{AMDAT} focuses on versatile and efficient implementation of analysis tools for characterizing structure and dynamics in equilibrium and non-equilibrium simulations of polymers, supercooled liquids, glass-forming materials, and other complex fluids.

\subsection{Why \textsf{AMDAT}?}

\textsf{AMDAT} is a molecular dynamics simulation trajectory analysis toolkit designed primarily for polymers, soft-matter, glass-forming liquids, and complex fluids.
Its design emphasizes in-memory handling of trajectories, efficient sampling over many decades of simulation time, and modular data structures that make it straightforward to assemble customized analysis workflows.
These choices, as follows, distinguish \textsf{AMDAT} from many existing analysis packages and make it well suited to computing the structural and dynamical observables commonly probed in studies of supercooled liquids and complex fluids.

\begin{enumerate}
    \item \textbf{In-memory trajectory handling for rapid analysis:}
    \textsf{AMDAT} reads the full molecular trajectory into memory upon initialization, greatly reducing file input/output operations and enabling efficient reuse of the same data across many analyses.
    While this approach increases memory requirements (a typical memory footprint is on the order of $2$--$3\times$ the trajectory file size), it allows near-instantaneous access to stored frames and significantly accelerates cross-frame calculations.
    As discussed below, an efficient time-sampling scheme helps keep this upfront memory cost tractable while still enabling analysis over long simulation times.
    This is particularly advantageous for two-time dynamical observables that compare configurations at different points along the trajectory, without repetitive data retrieval steps.

    \item \textbf{Exponential time sampling for long-timescale dynamics:}
    Standard operation of \textsf{AMDAT} involves reading and analyzing frames spaced exponentially in time.
    This strategy allows dense sampling at short times and progressively coarser sampling at longer times, enabling efficient analysis of molecular dynamics spanning many orders of magnitude in time without requiring prohibitively large trajectory files.
    We illustrate these benefits in the examples below in the section on the ``Benefits of Exponential Time Scaling.''
    
    \item \textbf{Modular and flexible data abstractions:}
    \textsf{AMDAT} implements data structures such as trajectory lists, neighbor lists, multibody lists, and value lists that can be combined, filtered, and manipulated to construct complex analyses.
    This modular approach makes it straightforward to tailor analyses to diverse molecular systems without modifying core code.
  
    \item \textbf{Extensive, tested suite of structural and dynamical observables:}
    \textsf{AMDAT} provides a broad set of well-tested routines, including calculations of structure factors, intermediate scattering functions, neighbor correlations, radial distribution functions, and a variety of other structural and dynamic correlation functions relevant to amorphous and glassy systems.
    These capabilities have been developed and maintained within our group for more than 15 years and have been used in dozens of published studies on polymers, supercooled liquids, and complex fluids~\cite{Kawak2024, Kawak2025, Kawak2025a,simmons_nature_2011,simmons_generalized_2012,lang_interfacial_2013,simmons_response_2013,mackura_enhancing_2014,marvin_nanoconfinement_2014,simmons_emerging_2016,mangalara_tuning_2015,lang_combined_2014,ruan_glass_2015,diaz-vela_temperature-independent_2018,merling_glass_2016,ruan_roles_2015,ye_understanding_2015,mangalara_three-layer_2016,cheng_design_2018,lee_correspondence_2017,smith_horizons_2017,mangalara_relationship_2017,mangalara_does_2017,chowdhury_spatially_2017,meenakshisundaram_designing_2017,hung_heterogeneous_2018,hung_string-like_2019,meenakshisundaram_design_2019,diaz_vela_microscopic_2020,hung_forecasting_2020,vela_probing_2020,rahman_near-substrate_2021,hao_mobility_2021,drayer_sequence_2022,ghanekarade_combined_2022,jaeger_temperature_2022,ghanekarade_nature_2021,tulsi_hierarchical_2022,ghanekarade_signature_2023, ghanekarade_glass_2023,hartley_interplay_2023,drayer_is_2024,xu_mixed_2024,hung_does_2025,yue_quantitatively_2024}.
    This long-term use and comparison with established results underpins our confidence that the routines yield reliable, publication-quality results across diverse molecular systems.

    \item \textbf{Scriptable workflows:}
    Analyses are specified via input scripts that support loops, conditionals, and variable evaluation, enabling reproducible and automated analysis pipelines.
    Outputs are written as plain-text, tab-delimited files or modified trajectory files compatible with \texttt{Python}, \texttt{Matlab}, \textsf{OVITO}, \textsf{VMD}, etc., such that analysis scripts can easily hand off to downstream tools.
    
    \item \textbf{Flexible file format support:}
    \textsf{AMDAT} reads standard and \texttt{custom} \textsf{LAMMPS} \texttt{dump}, standard \texttt{xyz}, and \textsf{GROMACS} \texttt{xtc} trajectory file formats.
    It can also be configured to read multiple per-atom properties for downstream analysis. Because of its object-oriented nature, read-in from other trajectory file styles can be readily implemented in future development.

\end{enumerate}

In the following sections, we describe \textsf{AMDAT}'s software design, discuss these features in greater depth, and illustrate the software's use with concrete examples and practical applications.


\section{Software Design and Implementation}


\subsection{Design Philosophy}

While applicable to a wide variety of simulated systems, \textsf{AMDAT} is designed to address the needs of researchers studying the structure and dynamics of non-crystalline materials such as supercooled liquids, polymers, and other glass-forming systems.
Accordingly, its design emphasizes computational efficiency for long-time, correlation-based observables, rather than visualization or graphical interactivity.

Building on the considerations outlined above, \textsf{AMDAT} adopts an in-memory, fast-compute model in which the trajectory is loaded into memory at runtime.
This reduces repeated file input/output and enables rapid access to particle coordinates and metadata across all stored time frames, which is particularly advantageous for analyses involving long-time correlations, exponential time spacing, or dynamic selection of particle subsets.
Although this choice increases memory usage, it substantially accelerates analysis workflows.

\textsf{AMDAT} is implemented in \texttt{C++} using an object-oriented, modular architecture.
Core classes represent the simulated system and its trajectories, while higher-level abstractions organize particles into reusable data objects that can be passed between analysis routines.
This separation between core data structures and analysis logic makes the code\-base easier to extend and to support the addition of new observables without intrusive changes to existing code.

A further guiding principle is that \textsf{AMDAT} is operated as a script\-able, non-interactive analysis engine.
Users interact with the software via a command-line interface and text-based input files, using a LAMMPS-like scripting language to specify system metadata, trajectory format, time scheme, selections, and analysis commands.
The input language supports basic control structures (loops, conditionals, variable evaluation), enabling complex workflows to be encoded in a transparent and reproducible manner while leaving visualization and post-processing to external tools better suited for those tasks.

\subsection{Core Architecture}

The architecture of \textsf{AMDAT} is built around a set of object-oriented data structures that enable flexible, composable analysis of molecular dynamics trajectories.
At the heart of this design is a hierarchy of classes that represent particles, molecules, and their trajectories, together with higher-level abstractions that organize these entities into reusable data objects for selection and analysis.

\subsubsection{System and Trajectory Classes}

The foundational class in \textsf{AMDAT} is the \texttt{System} object, which is responsible for reading trajectory files, storing metadata (e.g., box size, time scheme), and managing collections of particles and molecules.
Each particle or molecule is represented by a \texttt{Trajectory} object, which stores a time series of \texttt{Coordinate} instances.
Specialized subclasses such as \texttt{Atom\_Trajectory} and \texttt{Molecule} extend this base class to represent atomic and molecular entities, respectively. 

The \texttt{System} class also provides looping mechanisms over time frames, which are used by many analysis routines.
Loops over particles generally operate on user-defined subsets of particles via the \texttt{trajectory\_list} abstraction, described below.

\subsubsection{\texttt{trajectory\_list} and \texttt{trajectory\_bin\_list}}

The \texttt{trajectory\_list} is the central data object for analysis in \textsf{AMDAT}.
It stores a collection of particle trajectories, which may be static (fixed set of particles) or dynamic (membership varying with simulation frame).
Lists can be created explicitly (e.g., by designating particles of a given type or species) or generated by analysis routines (e.g., by selecting the most mobile particles at each time step).

For spatially resolved analyses, \textsf{AMDAT} provides the \texttt{trajectory\_bin\_list}, which partitions particles into spatial bins based on their coordinates.
Analyses can then be performed independently on each bin, enabling the study of interfacial gradients, local dynamics, and spatial heterogeneity.

\subsubsection{\texttt{multibody\_list}}

The \texttt{multibody\_list} abstraction represents groups of particles (e.g., molecules, functional groups, or particle clusters), the collective motion or internal structure of which are of interest.
\textsf{AMDAT} supports the creation of multibodies based on molecular connectivity, spatial proximity, or dynamic correlations (e.g., string-like motion).\cite{donati_stringlike_1998,pazmino_betancourt_string-like_2018,zhang_role_2015,betancourt_string_2014,zhang_string-like_2010,gebremichael_particle_2004,appignanesi_ballistic_2004,hung_does_2025,giovambattista_connection_2003}
Once defined, these multibodies can be analyzed for properties such as radius of gyration, orientational correlations, and reorientation dynamics.

\subsubsection{\texttt{value\_list} and \texttt{neighbor\_list}}

The \texttt{value\_list} is a versatile data object that stores scalar values for each particle at each time step.
These values can be computed during analysis (e.g., displacement magnitude) or read directly from trajectory files.
\texttt{value\_list} objects can be used to generate new \texttt{trajectory\_list} objects via thresholding or percentile selection, 
and can also be exported to \texttt{xyz} or \texttt{pdb} files for visualization (e.g., coloring particle by mobility or coordination number)~\cite{hartley_interplay_2023}.

The \texttt{neighbor\_list} is a specialized subclass of \texttt{value\_list} that tracks the number and identity of neighboring particles for each particle as a function of time.
Neighbor lists can be constructed using either distance-based cutoffs or Voronoi tessellation and are used in analyses such as neighbor decorrelation and persistent neighbor tracking. Pairs of particles can also be read in as a \texttt{neighbor\_list} to facilitate tracking of entities such as dynamic covalent bonds.

\subsubsection{Data Flow and Composability}

A key feature of \textsf{AMDAT}'s architecture is the composability of these data objects.
Many analysis routines take one or more data objects as input (e.g., a \texttt{trajectory\_list} or \texttt{multibody\_list}) and produce new objects as output (e.g., a \texttt{value\_list} or \texttt{trajectory\_bin\_list}).
These outputs can then be passed to subsequent analyses.
In this way, complex workflows can be constructed from simple components, and users can iteratively refine analysis pipelines by chaining together selections and observables.
For example, a \texttt{value\_list} containing per-particle displacements can be thresholded to generate a new \texttt{trajectory\_list} of ``mobile'' particles, which can then be analyzed for clustering or spatial correlations.

\subsection{Input and Control System}

\textsf{AMDAT} is operated via a command-line interface that reads a user-defined input file.
This input file serves as the central script for specifying the i) system to be analyzed, ii) trajectory location and format, iii) time scheme, iv) composition of constituents, and v) sequence of selections and analyses to be performed.
The design of the input system emphasizes transparency, reproducibility, and composability, enabling users to construct complex workflows using a straightforward, readable syntax.

\begin{lstlisting}[language=bash, basicstyle=\ttfamily\small, caption={Pseudocode structure for system, composition, selection, and analysis blocks in an \textsf{AMDAT} input file},label={alg:input}]
# --- SYSTEM BLOCK ---
<system_type>
<trajectory_file_type>
<filename1> [<filename2> $\ldots$]
<time_scheme>

# --- COMPOSITION BLOCK ---
<species1> <count1>   <species2> <count2>   $\ldots$   <speciesK> <countK> 
<type1>               <type2>               $\ldots$   <typeL>
<type1 in species1>   <type2 in species1>   $\ldots$   <typeL in species1> 
<type1 in species2>   <type2 in species2>   $\ldots$   <typeL in species2>
$\ldots$                   $\ldots$                   $\ldots$   $\ldots$
$\ldots$                   $\ldots$                   $\ldots$   $\ldots$
<type1 in speciesK>   <type2 in speciesK>   $\ldots$   <typeL in speciesK>

# --- SELECTION AND ANALYSIS BLOCKS ---
# These can appear in any order and quantity

# --- SELECTION BLOCK ---
# Create a list of atoms in species <species1>
create_list  s1_list
species  <species1>

# Create a list of atoms in species <species2> of type <type2>
create_list  s2t2_list
type_species  <species2>  <type2>

# Create a list of centers of mass for all molecules
create_multibodies all_mlist coms centroid all_molecule

# --- ANALYSIS BLOCK ---
msd msd_s1.dat
list s1_list

rdf data/rdf_s2t2_s1.dat asymmetric 100 -1 0
list s2t2_list
list s1_list

gyration_radius Rg_all.dat all_mlist

# --- MIXED BLOCK ---
create_distance_neighborlist my_neighbor_list 1.3 ./pair_distances.txt
\end{lstlisting}

\subsubsection{The System Block: System and Trajectory Specification}

The input file, illustrated schematically in Algorithm~\ref{alg:input}, begins with a block that defines the system type, trajectory file format, and associated metadata.
\textsf{AMDAT} supports both constant-volume (\texttt{<system\_type>}\allowbreak=\allowbreak\texttt{system\_nv}) and non-constant-volume
(\texttt{<system\_type>}\allowbreak=\allowbreak\texttt{system\_np}) systems, although some analyses are currently limited to the former.
Supported trajectory formats include \textsf{LAMMPS}-style \texttt{dump xyz} and \texttt{dump custom} files, as well as limited support for \textsf{GROMACS} \texttt{xtc} files.
The \texttt{custom\_manual} format allows users to map arbitrary columns in a trajectory file to values stored in a \texttt{value\_list} object, enabling flexible integration of auxiliary \emph{per-atom} data. This could include, for example, velocities or per-atom energies or forces.

The \texttt{<time\_scheme>} line specifies how time frames are sampled from the underlying trajectory.
Available schemes include \texttt{snapshot} (single-frame analysis), \texttt{linear} (uniform time spacing), and \texttt{exponential} (blocked, exponentially spaced frames).
The exponential scheme is particularly useful for time-correlation analyses spanning many orders of magnitude in time in glassy or supercooled systems (see the ``Benefits of Exponential Time Scaling'' section for further details).

\subsubsection{The Composition Block: Atom Types and Molecule Definitions}

Following the system block, the input file may contain a composition block describing the molecular makeup of the system.
The need for these lines depends on the specific \texttt{<trajectory\_file\_type>} used.
The first line lists the number of distinct species and their counts.
For example, a system with 15 polymer chains and 1000 solvent molecules would begin:
\texttt{polymer 15 solvent 1000}.
A header line then lists all atom types present in the trajectory, followed by one row per species that specifies how many atoms of each type appear in a single molecule of that species.
In the polymer/solvent example, the composition block shown in Algorithm~\ref{alg:input_eg1} assigns two atom types to each molecule:
\begin{lstlisting}[language=bash, basicstyle=\ttfamily\small, caption={Example composition block for a system with 15 polymer and 1000 solvent molecules.},label={alg:input_eg1}]
polymer 15  solvent 1000
1   2   3
10  5   0
0   1   2
\end{lstlisting}
Here, each polymer molecule contains 10, 5, and 0 atoms of types 1, 2, and 3, respectively.
Similarly, each one of the 1000 solvent molecules contains 0, 1, and 2 atoms of types 1, 2, and 3, respectively.

Users can define the composition in ways that best support their analysis.
For example, when molecular connectivity is unimportant, all atoms can be grouped into a single molecule. 

As another illustration, the ``Benefits of Exponential Time Scaling'' section presents three distinct composition specifications for a trajectory containing 20 atomistic polystyrene molecules.
\textsf{AMDAT} assumes that atoms are sorted in a consistent order across frames, does not currently support trajectories with varying compositions or particle counts in time, and does not natively read bonds, angles, or dihedrals.

\subsubsection{The Selection Block: Data Object Management}

After the system and composition are specified, the input file typically contains a sequence of commands that create and manipulate the core data objects defined in the Core Architecture section (\texttt{trajectory\_list}, \texttt{trajectory\_bin\_list}, \texttt{multibody\_list}, \texttt{value\_list}, and \texttt{neighbor\_list}).
For clarity, we refer to this part of the script as the selection block, although many commands play a dual role by both selecting particles and computing derived quantities (e.g., spatial binning creates a \texttt{trajectory\_list} of interfacial atoms), and in fact selection commands and analysis commands can alternate through the script.

Selection commands usually follow a simple pattern: a command that creates or modifies a named data object, followed by one or more lines that specify how that object is defined.
Lists support selection by atom index, molecule index, species type, atom type, or any combination of these.
For example, the two \texttt{create\_list} commands in Algorithm~\ref{alg:input} create two \texttt{trajectory\_list} objects: \texttt{s1\_list}, containing all atoms in \texttt{<species1>}, and \texttt{s2t2\_list}, containing atoms of type \texttt{<type2>} in \texttt{<species2>}.
Similarly, the \texttt{create\_multibodies} command constructs a \texttt{multibody\_list} in which each multibody is comprised of a specified set of particles.

\subsubsection{The Analysis Block: Scripted Analysis and Control Structures}

Once the desired data objects have been defined, analysis commands are used to compute observables based on them.
As illustrated in Algorithm~\ref{alg:input}, these commands usually operate on named data objects.
Most analyses consist of a first line that specifies the analysis type and output file, followed by one or more ``target'' lines that identify the data objects to be analyzed.

For example, the \texttt{msd} command in Algorithm~\ref{alg:input} sets the output file for mean-square displacement analysis via \texttt{msd msd\_s1.dat} and then specifies the target \texttt{trajectory\_list} via \texttt{list s1\_list}, which in this case contains all atoms in \texttt{<species1>} as defined by the composition block.
The \texttt{rdf} command exemplifies an analysis that operates on two lists, computing the radial distribution function for atoms in \texttt{s2t2\_list} around neighbors in \texttt{s1\_list}.
The \texttt{gyration\_radius} command targets the \texttt{multibody\_list} of molecule centroids and writes radius-of-gyration statistics to \texttt{Rg\_all.dat}.

Some commands both analyze existing data and create new data objects.
For instance, the \texttt{create\_\allowbreak distance\_\allowbreak neighborlist} command in Algorithm~\ref{alg:input} constructs a distance-based neighbor list with a cutoff of 1.3 and stores the resulting \texttt{neighbor\_list} as \texttt{my\_neighbor\_list} for use in subsequent analyses.

The scripting language supports a modest set of control structures, including \texttt{for} loops, \texttt{if...else} conditionals, and variable assignment via \texttt{constant} and \texttt{evaluate}.
These features allow users to automate repetitive tasks, perform parameter sweeps, and construct adaptive workflows.
Additional commands such as \texttt{print}, \texttt{wait}, and \texttt{user\_input} provide further flexibility for debugging and interactive execution.
\textsf{AMDAT} also accepts variables on the command line via the \texttt{-var} flag, enabling input scripts to incorporate externally specified parameters (e.g., running \textsf{AMDAT} with \texttt{/PATH/TO/AMDAT -i ./amdat.in -var cmd\_var 3} makes \texttt{\${cmd\_var}} available as a variable in \texttt{./amdat.in}).

\subsection{Extensibility and Development}

\textsf{AMDAT} is designed to be extensible, enabling researchers to implement new analysis methods, data structures, or input/output formats with minimal disruption to the existing codebase.
This is supported by a modular \texttt{C++} architecture that separates data representation from analysis logic and encourages the use of well-defined interfaces.

As described in the Core Architecture subsection, \textsf{AMDAT} distinguishes between a data-structure layer, which manages the simulated system and its trajectories, and an analysis layer, which operates on these data structures to compute observables and, where appropriate, create new data objects or output files.
Most analysis routines are implemented as classes that inherit from a common \texttt{Analysis} base class and operate on the standard data objects (\texttt{trajectory\_list}, \texttt{trajectory\_bin\_list}, \texttt{multibody\_list}, \texttt{value\_list}, \texttt{neighbor\_list}) defined in the Core Architecture section.
By interacting primarily with these abstractions, new analyses can be developed largely independently of the underlying trajectory format or system configuration.

For example, a new analysis class that computes a time-correlation function can be implemented to accept a \texttt{trajectory\_list} as input and produce a \texttt{value\_list} as output.
The resulting \texttt{value\_list} can then be passed to existing routines for thresholding, statistical analysis, or visualization.
Similarly, multi\-body-based analyses can be built on top of the \texttt{multibody\_list} abstraction to study collective motion or internal structural dynamics.

The \textsf{AMDAT} code\-base is hosted on GitHub\cite{Simmons_Amorphous_2025} (\url{https://github.com/dssimmons-codes/AMDAT}) and structured to facilitate collaborative development.
Developer-oriented documentation (\url{https://dssimmons-codes.github.io/AMDAT/}) outlines the key classes and interfaces required to implement new analyses, with each analysis class typically defined in its own header and source files following a consistent pattern for initialization, data access, and output generation.
Contributors are encouraged to follow the existing modular design and to build on the standard data objects described above.
Planned future improvements include expanded developer documentation and an importable Python interface.


\section{Representative Systems}
\label{sec:systems}


To demonstrate \textsf{AMDAT}'s versatility across diverse molecular contexts, we focus on six systems that span simple glass-forming liquids, reduced-dimensional models, coarse-grained polymers, polymer composites, and atomistic polymers.
Together, these systems provide a distinct set of structural and dynamical challenges and span a wide range of compositional and topological heterogeneity, length scales, and relaxation times, enabling us to showcase \textsf{AMDAT}'s breadth and capability to compute both static and dynamic observables for a variety of systems.

\subsection{Binary Lennard-Jones fluid (binLJ)}
This classic glass-forming mixture consists of two particle species interacting via Lennard-Jones (LJ) potentials.
We focus on the lowest equilibrium temperature reported in our prior works,\cite{hung_universal_2018,hung_string-like_2019} at a reduced Lennard-Jones temperature of $T^* = 0.3347$ and zero pressure.
At this state point, the system exhibits extremely slow dynamics, with a relaxation time of approximately $10^{5.94}\,\tau_{\mathrm{LJ}}$, among the longest equilibrium segmental relaxation times accessed in simulation.
The trajectory was generated in the NPT ensemble at $P=0$, with $N_1=6400$ particles of type 1 and $N_2=1600$ particles of type 2. Particles interact via the 12-6 binary Lennard Jones potential with energy and distance parameters $\varepsilon$ and $\sigma$. These parameters are both 1 for particles of type 1, are 0.50 and 0.88, respectively, for particles of type 2, and are 1.5 and 0.8, respectively, for cross-interactions. These conditions and interactions yield a number density of $\sim 1.17$.

\subsection{Two-dimensional binary Lennard-Jones fluid (binLJ2D)}
This system mirrors the composition and interaction scheme of binLJ but is confined to two dimensions, providing a test of \textsf{AMDAT}'s ability to handle reduced dimensionality.
Simulated under conditions analogous to binLJ, this model highlights how structural and dynamical signatures evolve when dimensional constraints are imposed.

\subsection{Kremer–Grest polymer chain (KG)}
A widely used coarse-grained bead-spring polymer model, the KG system captures essential features of polymer dynamics while remaining computationally tractable.
Here, we analyze the lowest temperature reported in our previous studies,\cite{hung_universal_2018,hung_string-like_2019} at $T^* = 0.3854$, where the relaxation time reaches $10^{6.88}\,\tau_{\mathrm{LJ}}$, placing it among the most sluggish equilibrium states probed for any simulated glass-former.
The system comprises 400 chains with 20 monomers per chain, simulated in the NPT ensemble at $P=0$.

\subsection{Nanoparticle Filled End-Crosslinked KG (PNC)}
This system is used in our recent work to study the mechanical deformation of filled elastomers~\cite{Smith2017, Smith2019, Kawak2024, Kawak2025, Kawak2025a} and enables illustration of how \textsf{AMDAT} analyzes heterogeneous systems.
It contains 5000 chains with 20 KG beads each, 2500 crosslinker beads, and 50 distinct clusters of 7 icosahedral particles with each particle containing 147 KG beads (center bead with three surrounding shells).
The KG chains contain two chain ends that crosslink to up to 2 crosslinker beads, while crosslinker beads have a functionality of 4.
The system is crosslinked to 95\% of all possible crosslinks, forming a well-formed elastomer network.
Filler clusters are dispersed throughout the elastomer.
Finally, to show how attractive nanoparticle surfaces impact the dynamics of neighboring polymer segments, the polymer--filler LJ interaction strength is three times stronger than self interactions.
This trajectory was prepared using an MD simulation in the NPT ensemble at a unit reduced temperature and zero pressure.
For more details on the preparation protocol, we invite the reader to review the methods section of our earlier publication~\cite{Kawak2025a}.

\subsection{Atomistic polystyrene 30-mer (PS-30mer)}
This all-atom representation of polystyrene (PS) employs the OPLS force field to simulate atomistic polystyrene as described in our prior publications\cite{hung_universal_2018,hung_forecasting_2020}
and provides an exemplar for \textsf{AMDAT}'s ability to handle chemically detailed systems.
The trajectory corresponds to 30 monomers per chain and 13978 total atoms, simulated in the NPT ensemble at $T=483$ K.

\subsection{Atomistic polystyrene 100-mer (PS-100mer)}
This all-atom representation of PS is identical to the system above but contains 100 monomers of styrene 
(32040 total atoms) and was simulated at $T=478$ K.


\section{Capabilities and Features}
\label{sec:capabilities}


In this section, we summarize the main classes of observables that \textsf{AMDAT} computes, illustrated for the representative systems described above.

\subsection{System-Average Static Structural Properties}

\textsf{AMDAT} computes a range of real-space and reciprocal-space structural observables that characterize packing, composition, and molecular shape.
Unless otherwise noted, these analyses operate on user-defined \texttt{trajectory\_list} or \texttt{trajectory\_bin\_list} objects.

\paragraph{Pair structure and composition.}
Real-space structure is quantified via the radial distribution function $g(r)$ and the related radial neighbor count, a non-volume-normalized measure of neighbors as a function of distance (commands \texttt{rdf} and \texttt{rnf}, respectively).
Composition analysis (\texttt{composition}, \texttt{composition\_vs\_time}) provides counts of particles by species or type, either as time-averaged system-wide statistics or, when applied to binned trajectories, as spatially resolved density or composition profiles.

\paragraph{Clustering and connectivity.}
Distance-based cluster identification (\texttt{clustered\_list}) can be used to determine which particles satisfy a specified clustering criterion.
The resulting lists can then be analyzed with the composition tools to quantify, for example, cluster size distributions or the fraction of particles in clusters of a given type.

\paragraph{Reciprocal-space structure.}
The static structure factor $S(q)$ can be computed in two ways.
The \texttt{structure\_factor} command evaluates $S(q)$ via the inverse-density approach using all particles in the simulation box (currently limited to fixed-dimension cubic boxes), while \texttt{structure\_factor\_from\_rdf} performs a radially symmetric discrete Fourier transform of a previously computed $g(r)$, yielding an $S(q)$ that is applicable in more general geometries at the cost of somewhat lower statistical strength.

\paragraph{Molecular shape and orientational order.}
For molecular or multibody entities, \textsf{AMDAT} provides system- and time-averaged measures of shape and orientation.
The \texttt{gyration\_radius} command, applied to a \texttt{multibody\_list}, computes radii of gyration for molecules or clusters.
Orientational correlations with respect to an externally defined vector can be evaluated via the \texttt{orientational\_correlation} command, enabling, for example, quantification of alignment relative to a field or interface normal.

\begin{figure}[!ht]
    \centering
    \includegraphics[]{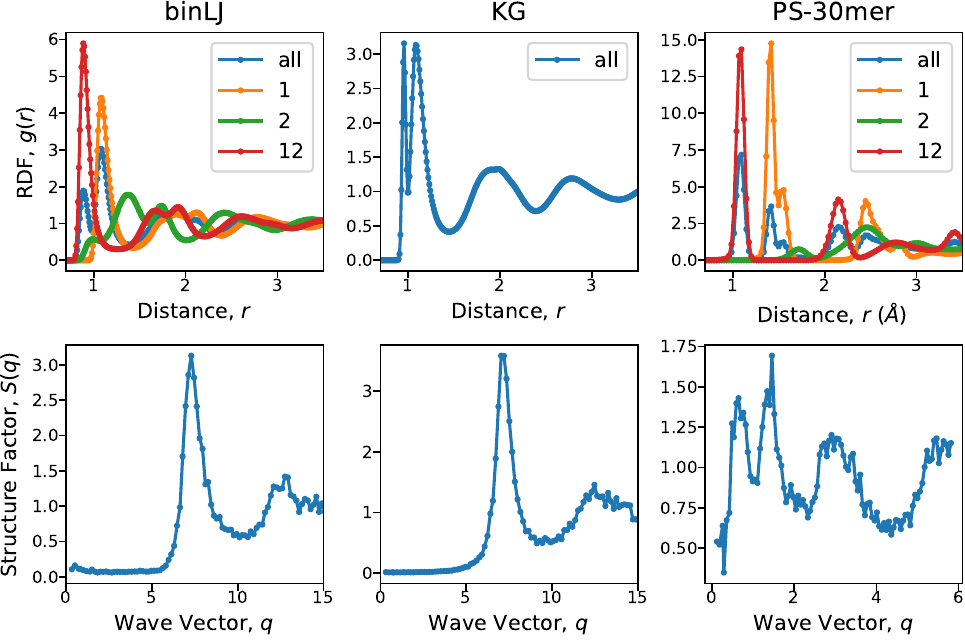}
    \caption{Radial distribution functions (top) and structure factors (bottom) for binLJ, KG, and PS-30mer systems from left to right.}
    \label{fig:static}
\end{figure}

Figure~\ref{fig:static} illustrates several of these static observables for representative systems.
For binLJ and binLJ2D, radial distribution functions are shown for all particles and for individual species, including asymmetric RDFs for cross interactions.
For PS-30mer, RDFs are decomposed by carbon and hydrogen atom types.
Structure factors $S(q)$ are computed for all particles in each system.

\subsection{System-Average Dynamical Properties}

\textsf{AMDAT} provides a suite of tools for probing translational, rotational, and neighbor-shell dynamics, including both widely used measures and observables tailored to glass-forming liquids and polymers.
Unless otherwise noted, these analyses operate on user-defined \texttt{trajectory\_list} objects.

\paragraph{Single-particle translational dynamics.}
The mean-square displacement (MSD) is computed via the \texttt{msd} command, with options for three-dimensional displacements or projections onto a two-dimensional plane.
Complementary quantities include the mean displacement and its incremental form as functions of time (\texttt{mean\_displacement}, \texttt{incremental\_mean\allowbreak\_\allowbreak displacement}), which can also be evaluated on binned trajectories (\texttt{trajectory\_bin\_list}) for spatially resolved analyses.
The non-Gaussian parameter (\texttt{ngp}) and particle displacement distribution (\texttt{displacement\_dist}) quantify deviations from diffusive, Gaussian behavior.
Distributions of Debye–Waller factors and their inverse ``stiffness'' values (\texttt{u2\_dist} and \texttt{stiffness\_dist}, respectively) provide additional measures of caging and local vibrational confinement.

\paragraph{Scattering and correlation functions.}
In reciprocal space, the self part of the intermediate scattering function $F_s(q,t)$ is computed via the \texttt{isfs} command over user-specified ranges of wavevectors.
The full intermediate scattering function is computed via the \texttt{isf\_list} command, although this is currently only fully implemented for fixed-size, cubic systems due to a shared underlying architecture with the inverse space structure factor calculation.
In real space, the self, distinct, and total parts of the Van Hove correlation function $G(r,t)$ are obtained using the \texttt{vhs}, \texttt{vhd}, and \texttt{vht} commands, respectively.
Reorientational autocorrelation functions for molecular or multibody vectors are computed via the \texttt{raf} command applied to a \texttt{multibody\_list}, enabling analysis of rotational relaxation.

\paragraph{Neighbor-shell dynamics.}
Neighbor-shell dynamics are characterized by the neighbor decorrelation function, computed with \texttt{neighbor\_decorrelation\_function} from a preconstructed \texttt{neighbor\_list} and a corresponding \texttt{trajectory\_list}.
This observable monitors the persistence and turnover of local coordination shells in time.

\begin{figure}[!ht]
    \centering
    \includegraphics[width=6.5in]{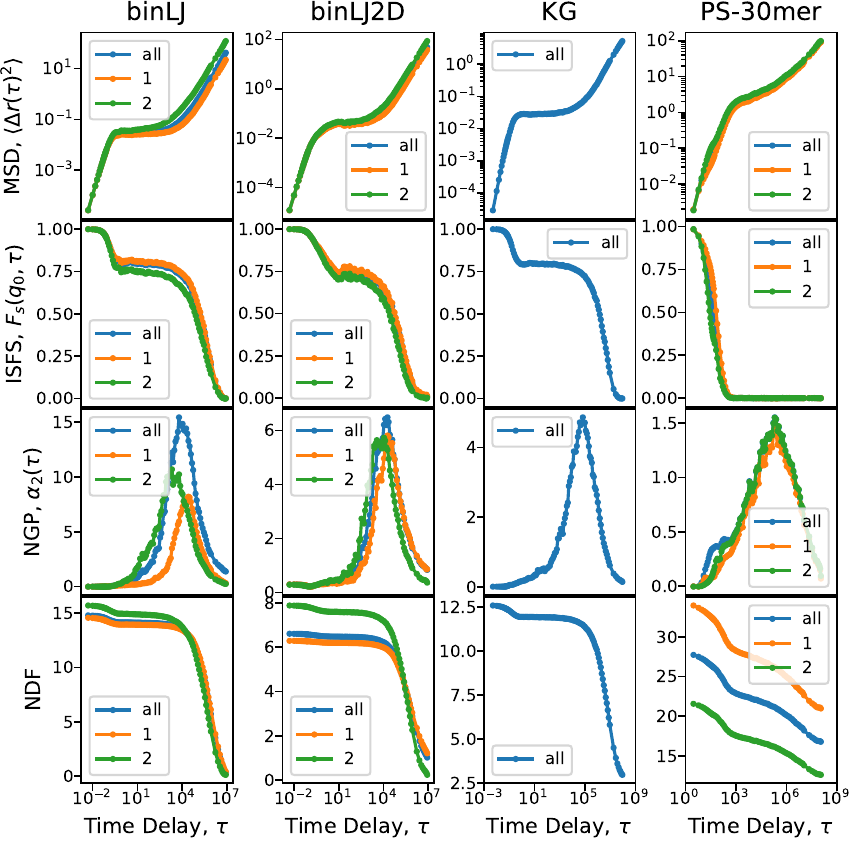}
    \caption{Dynamical properties (MSD, ISFS, NGP, NDF) for binLJ, binLJ2D, KG, and PS-30mer systems from left to right.}
    \label{fig:dynamic}
\end{figure}

\begin{figure}[!ht]
    \centering
    \includegraphics[]{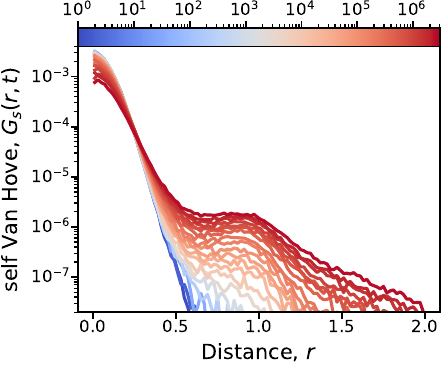}
    \caption{Self part of the Van Hove correlation function for the KG system. The figure shows isochronous curves versus distance colored by time according to the displayed color bar.}
    \label{fig:vhs-KG}
\end{figure}

Figure~\ref{fig:dynamic} illustrates several of these time-dependent properties with species-level resolution for the representative systems.
Mean-square displacement, the self part of the intermediate scattering function, the non-Gaussian parameter, and the neighbor decorrelation function are shown for binLJ, binLJ2D, KG, and PS-30mer.
Figure~\ref{fig:vhs-KG} shows the self part of the Van Hove correlation function for the KG system as isochronous curves of $G_s(r,t)$ versus distance, colored by time.

\subsection{Local and Per-Bead Analyses}

All of the metrics presented thus far are averaged properties across the entire system or on a subset of atoms defined by species or type.
\textsf{AMDAT} also supports computation of per-particle quantities that can be used for spatially resolved analysis, thresholding, or external visualization.

Several analysis routines produce per-atom fields stored in \texttt{value\_list} objects (e.g., displacements, local order parameters, or coordination numbers).
These \texttt{value\_list}s can be exported alongside atomic coordinates, for example as the $\beta$ column of a \texttt{pdb} file, and visualized with molecular visualization software such as \textsf{VM} or \textsf{OVITO}.

\begin{figure}[!ht]
    \centering
    \includegraphics[]{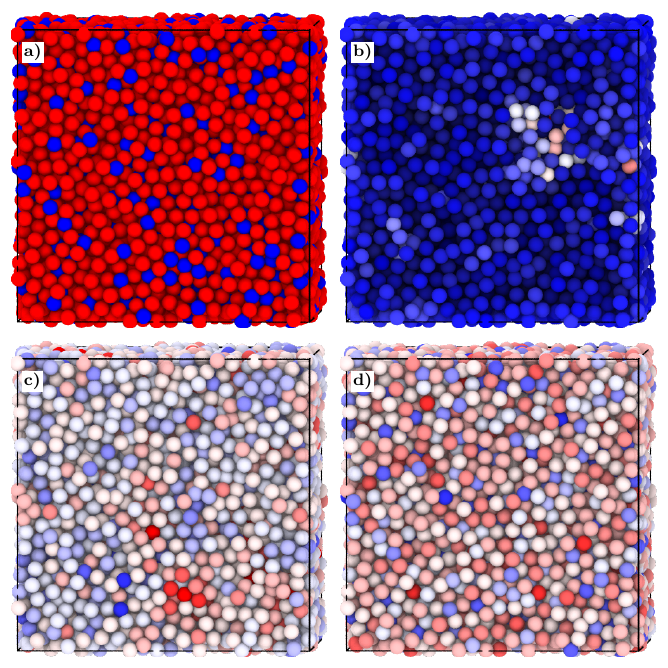}
    \caption{Configurations of the binary Lennard-Jones system, colored by different per-particle properties.
    Panels show a) atom type, b) displacement after 1211.42 $\tau_{LJ}$, c) number of neighbors within a 1.4 $\sigma_{LJ}$ cutoff, and d) number of neighbors in a Voronoi-tessellated cell.
    Visualization was performed using \textsf{OVITO}~\cite{OVITO}.}
    \label{fig:binLJ_per_bead}
\end{figure}

\begin{figure}[!ht]
    \centering
    \includegraphics[]{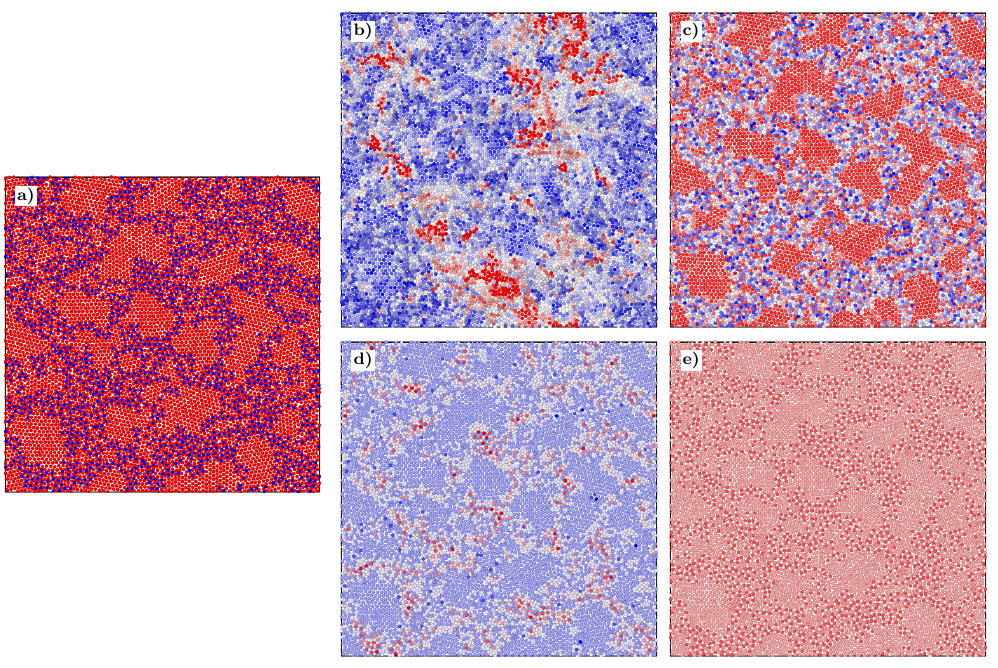}
    \caption{Configurations of the 2D binary Lennard-Jones system, colored by different per-particle properties.
    Panels show a) atom type, b) displacement after 1211.42 $\tau_{LJ}$, c) 2D (xy) 6-fold hexatic order parameter, d) number of neighbors within a 1.4 $\sigma_{LJ}$ cutoff, and e) number of neighbors in a Voronoi-tessellated box.
    Visualization was performed using \textsf{OVITO}~\cite{OVITO}.}
    \label{fig:binLJ2D_per_bead}
\end{figure}

Figures~\ref{fig:binLJ_per_bead} and~\ref{fig:binLJ2D_per_bead} illustrate such per-particle visualizations for the 3D and 2D binary LJ systems, respectively.
In both figures, all panels show the same snapshot colored by different per-particle properties, with panel (a) indicating the two atom types.

Panel (b) colors particles by their displacement over a specified time interval in the exponential time scheme.
This field is generated by the \texttt{displacement\_list} command, which constructs a \texttt{value\_list} containing per-particle displacements at a selected time delay.
The resulting \texttt{value\_list} can be written to a \texttt{pdb} file using \texttt{value\_list write\_pdb}, enabling direct visualization of mobile and immobile regions.

Panels (c)–(e) demonstrate additional per-particle fields constructed within \textsf{AMDAT}.
The \texttt{nfold} command computes the 2D 6-fold hexatic order parameter,\cite{tanaka_critical-like_2010} while \texttt{create\_distance\allowbreak\_\allowbreak neighborlist} and \texttt{create\_voronoi\_neighborlist} generate neighbor lists from which coordination numbers are obtained based on distance cutoffs or Voronoi tessellation, respectively.
Each of these quantities is stored in a \texttt{value\_list} and exported via \texttt{value\_list write\_pdb} to produce the corresponding visualizations.

Together, these examples highlight how \textsf{AMDAT}'s per-particle analyses and \texttt{value\_list}-based workflows can be combined with external visualization tools to reveal spatial heterogeneity in structure, dynamics, and local environment.

\subsection{Neighbor- and Multibody-Based Metrics}

Beyond system-averaged observables and per-particle fields, \textsf{AMDAT} provides tools for characterizing local environments, multibody structure, and cooperative motion.
These quantities are built on the neighbor- and multibody-based data objects described in the Core Architecture section.

\paragraph{Neighbor statistics and decorrelation.}
Neighbor lists can be constructed using either distance-based cutoffs or Voronoi tessellation via the \texttt{create\_distance\_neighborlist} and \texttt{create\_voronoi\_neighborlist} commands.
The resulting \texttt{neighbor\_list} objects encode both the number and identity of neighbors for each particle as a function of time and can be used to compute static and dynamic measures of local environment.

\begin{figure}[!ht]
    \centering
    \includegraphics[]{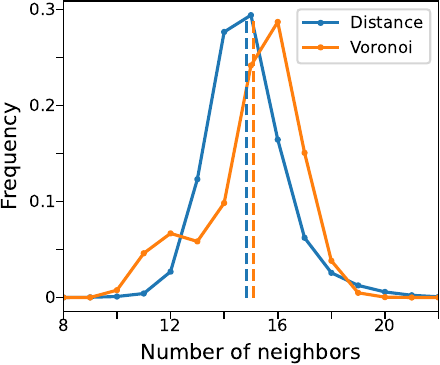}
    \caption{Histogram of number of neighbors for the binLJ system from two neighbors lists constructed using distance criteria (1.4$\sigma_{LJ}$) and Voronoi tessellation.
    Dashed lines indicate the average coordination number for each distribution.}
    \label{fig:neighs-binLJ}
\end{figure}

Static neighbor statistics, such as distributions of coordination number, can be obtained directly from a \texttt{neighbor\_list}.
Figure~\ref{fig:neighs-binLJ} shows an example for the binLJ system, comparing histograms of neighbor counts obtained from distance-based and Voronoi-based neighbor definitions.

Temporal evolution of local environments is quantified by the neighbor decorrelation function, computed via the \texttt{neighbor\_decorrelation\_function} command using a \texttt{neighbor\_list} together with a corresponding \texttt{trajectory\_list}.
This measure monitors the persistence and turnover of each particle's neighbor shell and is particularly useful for studying cage breaking and cooperative rearrangements in glass-forming systems.
Persistent neighbors and their clustering can be further analyzed through the \texttt{persistent\_neighbors} command, which constructs multibodies from particles that remain neighbors over user-specified time intervals.

\paragraph{Multibody size, shape, and cooperative motion.}
For molecular or cluster-level analyses, \textsf{AMDAT} builds on \texttt{multibody\_list} objects that group particles into molecules, clusters, or dynamically defined structures.
Several analyses characterize these multibodies:
\begin{itemize}
    \item \texttt{gyration\_radius}: Computes radius-of-gyration statistics for multibodies (e.g., polymer chains, clusters).
    \item \texttt{size\_statistics}: Reports size distributions of multibodies, such as cluster size distributions in systems with dynamic aggregation.
    \item \texttt{orientational\_correlation}, \texttt{raf}: Evaluate orientational and reorientational correlation functions for vectors defined within each multibody, enabling analysis of rotational relaxation and alignment.
    \item \texttt{string\_multibodies}, \texttt{clustered\_list}: Identify and analyze collective motion and clustering, for example by constructing multibodies corresponding to string-like motions or spatial clusters of highly mobile particles.
\end{itemize}

These neighbor- and multibody-based metrics complement global observables by resolving the structure and dynamics of local environments and cooperative rearrangements.

\subsection{Advanced Workflows}

Many of \textsf{AMDAT}'s capabilities are designed to be combined into flexible analysis pipelines.
In addition to the specialized commands above, several features facilitate more advanced or system-specific workflows.

\paragraph{Custom inputs, local resolution, and dynamic subsets.}
The \texttt{custom\_manual} trajectory format allows users to map arbitrary columns of a trajectory file to \texttt{value\_list} objects, making it straightforward to incorporate auxiliary per-atom data (e.g., local stress, order parameters, or externally computed fields) into subsequent analyses.

Spatial resolution can be introduced via \texttt{trajectory\_bin\_list} objects, which support binning along Cartesian axes or relative to reference particles or molecules.
Many observables described above (e.g., composition, MSD, or local structure metrics) can be computed on these binned lists to obtain profiles across interfaces, gradients, or confinement geometries.

Intricate selections and dynamic subsets are supported by combining selection commands with per-particle fields.
For example, asymmetric radial distribution functions and dynamic properties can be computed for specific atom types, molecular substructures, or subsets defined by mobility thresholds.
Lists of the fastest or slowest particles over a given time interval can be identified from displacement-based \texttt{value\_list}s and then analyzed for clustering, local environment, or multibody statistics.

\paragraph{Generic auto- and cross-correlation of \texttt{value\_list}s.}
Many of the decorrelation and correlation functions described earlier (e.g., neighbor decorrelation, reorientational relaxation, displacement-based statistics) can be viewed as specific cases of a more general framework for correlating scalar fields in time.
\textsf{AMDAT} exposes this framework through two generic commands:
\begin{itemize}
    \item \texttt{autocorrelate\_value\_list}: Computes time autocorrelation functions of a single \texttt{value\allowbreak\_\allowbreak list}, enabling users to define and analyze arbitrary time-correlation functions of per-particle scalar quantities.
    \item \texttt{crosscorrelate\_value\_lists}: Computes cross-correlation functions between two \texttt{value\allowbreak\_\allowbreak list} objects, allowing, for example, correlations between local mobility and local structure, or between two independently computed fields.
\end{itemize}

Any property that can be stored in a \texttt{value\_list}---whether read in from a trajectory file, constructed from neighbor information, or generated by previous analyses---can be passed to these routines.
In practice, this provides a recipe for defining and computing a wide range of time correlation and cross-correlation functions without writing new analysis code, making \textsf{AMDAT} a flexible platform for user-defined correlation analyses.


\section{Examples, Use Cases, and Applications}


\subsection{Benefits of Exponential Time Scaling} \label{sec:sec:exp}

For glass-forming liquids and polymers with slow relaxation, exponential time blocking provides markedly better statistical efficiency and time-range coverage than uniform (linear) output.
In a linear scheme with $T$ frames spaced by $\Delta n$ integration steps, the number of frame pairs available at delay $\tau$ decreases as $S(\tau)=T-\tau/\Delta n$, leading to excessive redundancy at short delays and poor statistics at long delays.
By contrast, blocked exponential spacing keeps a fixed number of start times per block for all delays within a block, yielding more uniform statistical quality across multiple decades of time while requiring fewer total pair evaluations.%

The exponential blocking scheme also limits the analysis to a single trajectory.
Without this scheme, the standard is to save multiple trajectories with different spacings to permit analysis of multiple decades of dynamics.
To illustrate this with an example, if a researcher is interested in 8 decades from $10^{-3}$ to $10^5$ picoseconds, a single trajectory would require a spacing of $10^{-3}$ picoseconds and result in $10^8$ snapshots!
Therefore, a linear scheme necessitates multiple trajectories with different spacings to appropriately span multiple decades in a tractable manner.
In contrast, an exponential scheme, with a small initial spacing increasing exponentially, has all of the time frame pairs necessary to analyze dynamics across multiple decades in time spacings.

Below we show the minimal differences in \textsf{LAMMPS} input scripts that produce (i) a linearly spaced trajectory (Algorithm ~\ref{alg:lmp_linear}) and (ii) an exponentially blocked trajectory (Algorithm~\ref{alg:lmp_exp}).
We do so for the PS-100mer system, where the integration time step is 1 picosecond (ps).
The linear script dumps every fixed number of steps; the exponential script computes the \emph{next} output time on-the-fly using a base $b$, a per-block size $K$, and a number of blocks $I$.
The corresponding \texttt{<time\_scheme>} line in the \textsf{AMDAT} input must match the way frames were written using \textsf{LAMMPS}.
\begin{lstlisting}[language=bash, basicstyle=\ttfamily\small, caption={\textsf{LAMMPS} input (linear spacing): output every 13{,}529 MD steps over a fixed run.}, label={alg:lmp_linear}]
# --- dump evenly spaced (linear) trajectory frames ---
dump linear all custom 13529 ./linear.traj type x y z ix iy iz
dump_modify linear first yes append yes sort id
\end{lstlisting}
\noindent
In this example, the dump stride is $13{,}529$ ps.
This number was chosen to yield 771 frames, equal to the number of frames in the exponential scheme below.
Given that the syntax for the \texttt{linear} time scheme in \textsf{AMDAT} is \texttt{linear <number\_of\_snapshots> <time\_spacing>}, the corresponding \texttt{<time\_scheme>} line in the \textsf{AMDAT} input script for this situation is \texttt{linear 771 13600}.

Linear spacing is in general a poor way to study dynamical correlations.
For each time delay, dynamical correlations use all (or a subset of all) frames that are separated by that time delay.
For any time delay, $\Delta t$, the number of frames available for correlation computes is $S(\Delta t) = T - \frac{\Delta t}{\Delta \tau}$, where $\Delta \tau$ is the even spacing used (13529 ps above) and $T$ is the total number of frames.
This has two important consequences: small separations are oversampled with many possible frame pairs, while large time delays feature a few possible pairs of frames.
The former problem results in a large analysis time for minimal additional statistical accuracy.
To illustrate this problem with the example above, for separations of $10^5$, $10^6$, and $10^7$, there are 762, 696, and 30 total pairs, respectively, for correlation analysis.
The latter problem of large time delays is exacerbated by the fact that possible unique pairs are correlated (e.g., with $T=1000$, $\Delta \tau=10^3$, time delay $\Delta t=997 \times10^3$ has $S=3$ possible pairs for analysis but they are frames (0, 997), (1, 998), and (2, 999), resulting in highly correlated results).

Next, we present the modifications necessary in the \textsf{LAMMPS} script to provide an exponentially spaced trajectory for analysis using \textsf{AMDAT} in Algorithm~\ref{alg:lmp_exp}.
\begin{lstlisting}[language=bash, basicstyle=\ttfamily\small, caption={\textsf{LAMMPS} input (exponential blocking): compute the next output step via variables (base b, block size K, blocks I).}, label={alg:lmp_exp}]
dump exp all custom 1 ./exp.traj type x y z ix iy iz
dump_modify exp append yes first yes sort id

# --- variables controlling exponential spacing ---
# Expected to be set upstream or via command-line -var flag:
#   exp_base = b, blocksize = K, blocknumber = I
# Example: lmp -var exp_base 1.2 -var blocksize 77 -var blocknumber 10

thermo_modify line yaml
thermo_modify flush yes

variable lastinner     equal "floor(v_exp_base^(v_blocksize-1))"
variable lastouter     equal "v_lastinner * v_blocknumber"
variable getstep       equal step
variable innerstep     equal "v_getstep % v_lastinner"
variable blockindex    equal "floor(v_getstep / v_lastinner)"

# logic for log spacing within each block
variable nextlinear        equal "v_innerstep + 1"
variable nextlogindex      equal "1 + ceil(ln(v_nextlinear)/ln(v_exp_base))"
variable nextinnerindex    equal "ternary(v_nextlinear < v_nextlogindex, v_nextlinear, v_nextlogindex)"
variable nextloginnerstep  equal "floor(v_exp_base^(v_nextinnerindex - 1))"
variable nextinnerstep     equal "ternary(v_nextinnerindex > v_nextloginnerstep, v_nextinnerindex, v_nextloginnerstep)"
variable nextoutput        equal "v_nextinnerstep + v_blockindex * v_lastinner"

thermo          v_nextoutput
dump_modify     exp every v_nextoutput

# total integration steps = last step in a block * number of blocks
run v_lastouter
\end{lstlisting}
This script outputs frames at times that are exponentially spaced, repeated across $I=10$ blocks; the total run length is the last time in one block times the number of blocks (\texttt{v\_lastouter}$=1041776\times10$).
The corresponding \texttt{<time\_scheme>} line in the \textsf{AMDAT} input script is \texttt{exponential 10 77 1.2 0 0 1.0}.

To postprocess the resulting trajectories using \textsf{AMDAT}, we utilize the script in Algorithm~\ref{alg:amdat_exp}.
\begin{lstlisting}[
  basicstyle=\ttfamily\small,
  caption={\textsf{AMDAT} input for reading an exponentially blocked trajectory containing 20 chains of 100 styrene monomers},
  label={alg:amdat_exp}
]
system_np
custom
./exp.traj
exponential 10 77 1.2 0 0 1.0
polymer 20
1 2 3 4 5 6 7 8 9
99 99 500 500 100 1 100 3 200

#create a list of all atoms in the system
create_list all 
all

#compute the mean square displacement of all particles as a function of time and save to a file named msd.dat
msd ./msd.dat
list all
\end{lstlisting}
In Algorithm~\ref{alg:amdat_exp}, we choose \texttt{system\_np} and \texttt{custom} to match the \texttt{fix npt} and \texttt{dump custom} commands in \textsf{LAMMPS}.
After the trajectory name, we outline the exponential block parameterization.
Specifically, \texttt{exponential $I$ $K$ $b$ $\mathrm{frt}$ $a_0$ $\Delta \tau$} in \textsf{AMDAT} maps to \textsf{LAMMPS} variables \texttt{blocknumber} ($I$), \texttt{blocksize} ($K$), and \texttt{exp\_base} ($b$).
Here we use $\mathrm{frt}=0$ (treat the last frame of a block as the first of the next for pairing), $a_0=0$ (first exponent index), and $\Delta \tau=1.0$ ps.
Unlike linearly spaced trajectories, $\Delta \tau$ for this scheme is the \texttt{timestep} employed in \textsf{LAMMPS}.
For the linear trajectory, we replace the exponential line with \texttt{linear $T$ $\Delta t$}.
In \textsf{AMDAT}, this corresponds to $T$ frames separated by $\Delta \tau$ integration steps (in time units) in \textsf{LAMMPS} (here, $T=771$, $\Delta t=13529$).

Next, the ``composition block'' lines tell \textsf{AMDAT} what particles to expect and how to define molecules and their composition for analysis.
In the above example in Algorithm~\ref{alg:amdat_exp}, we define 20 polymers with the name \texttt{polymer}, with 9 atom types (1 to 9) with their counts \emph{per polymer chain}.
There are in general many ways to define the composition block for any trajectory.
For example, if one is not interested in per chain behavior, the entire composition can be defined as a single molecule (Algorithm~\ref{alg:composition_2}). 

Finally, we create a list of all atoms using \texttt{create\_list all\char`\\n all} and calculate the mean-square displacement using the \textsf{AMDAT} \texttt{msd} command.

\begin{lstlisting}[basicstyle=\ttfamily\small, caption=Alternative for \textsf{AMDAT} composition block in Algorithm~\ref{alg:amdat_exp} with all atom types in a single molecule, label={alg:composition_2}]
all 1
1 2 3 4 5 6 7 8 9
1980 1980 10000 10000 2000 20 2000 60 4000
\end{lstlisting}

\begin{figure}[!ht]
     \centering
     \includegraphics[]{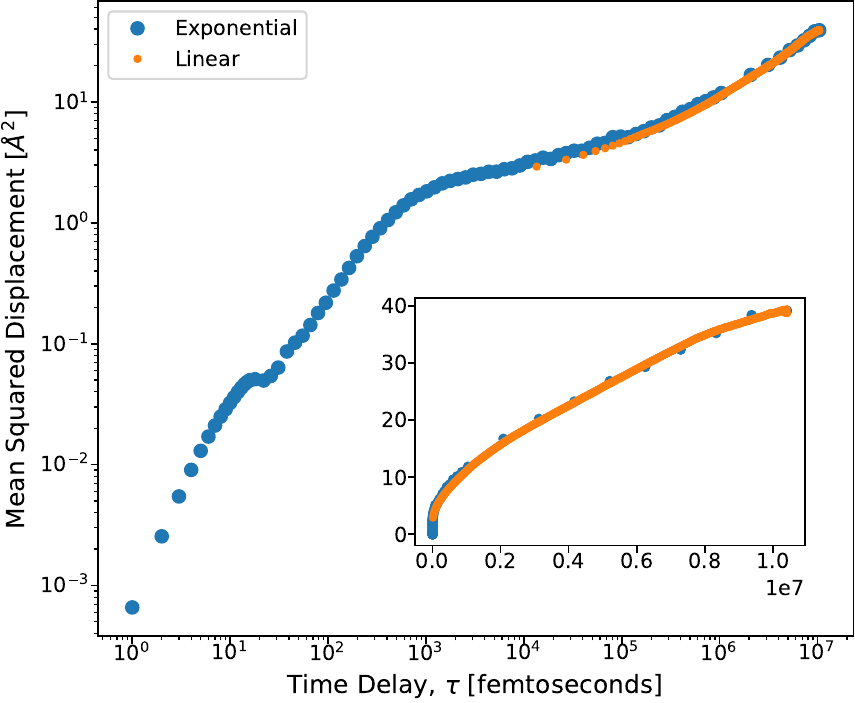}
     \caption{MSD comparison between linear and exponential time spaced trajectories from the same MD simulation of PS-100mer.}
     \label{fig:exp-vs-linear}
 \end{figure}

In Fig.~\ref{fig:exp-vs-linear}, we present this analysis for the PS-100mer system with the linear and exponential time schemes.
Despite employing the same number of frames in both trajectories, the exponential scheme probes more than double the logarithmic time span of the linear scheme.
The exponential scheme achieves (a) broader time coverage, (b) higher short-time resolution, and (c) markedly fewer time-pair evaluations for comparable or better uncertainty at long delays, compared with linear spacing at the same total run length.
In practice, this yields faster execution and improved statistical power for MSD and other correlation functions across the full time window.

\subsection{Local Resolution and Per-Atom Properties in a Model Polymer Nanocomposite (PNC)} \label{sec:local_pnc}

Polymer nanocomposites (PNCs) exhibit spatially heterogeneous properties due to the presence of rigid fillers and the effects of polymer--filler attractions.
In such systems, polymer segments near filler surfaces often display markedly slower dynamics than those in the bulk.
Here we demonstrate some of \textsf{AMDAT}'s capabilities spatially resolved analysis on an end-crosslinked Kremer–Grest (KG) network dispersed with carbon-black-like clusters, and show how to reproduce the workflow using \textsf{AMDAT} commands.

The PNC comprises end-crosslinked KG chains, with 5000 chains of strand length 20 and 2500 crosslinker particles; approximately 95\% of all possible end crosslinks are formed.
Rigid filler clusters mimic carbon black with each cluster consisting of 7 primary icosahedral particles.
Each icosahedral particle contains 147 beads starting with a center bead, inner shell, middle shell, and outer shell.
We load the elastomer with 50 such clusters or 26.1\% filler by volume.
This system is similar to those studied previously~\cite{Kawak2024, Kawak2025, Kawak2025a}.
Critically, the polymer--filler cross-interaction strength is set to 3$\times$ the polymer (and filler) self-interaction strength, ensuring a strong interphase where chains are substantially slowed relative to the bulk polymer.

\subsubsection{Locally resolved ISFS}

We use a \textsf{LAMMPS} simulation to output an exponentially spaced trajectory (Algorithm~\ref{alg:lmp_exp}) with a single block ($I{=}1$), block size $K{=}103$, base $b{=}1.2$, and base time unit $\Delta \tau{=}0.001$ (\textsf{AMDAT} \texttt{time\_scheme} line: \texttt{exponential 1 103 1.2 0 0 0.001}).
Given this setup, the \textsf{AMDAT} script begins with Algorithm~\ref{alg:amdat:pnc_isfs_1}.
\begin{lstlisting}[basicstyle=\ttfamily\small, caption={\textsf{AMDAT} PNC ISFS analysis part 1: System and composition blocks for PNC system}, label={alg:amdat:pnc_isfs_1}]
system_np
custom
./exp.traj
exponential 1 103 1.2 0 0 0.001
polymer 1 xlinkr 1 filler 50
1 2 3 4 5 6 7 8 9
90000 500 0 0 0 0 0 9500 0
0 0 326 0 0 0 0 0 2174
0 0 0 644 294 84 7 0 0
\end{lstlisting}
In the above setup, particle types 1, 2, and 8 are polymer internal bead, polymer end bead with less than two bonded crosslinkers, and polymer end bead with two bonded crosslinkers, respectively.
Further, particle types 3 and 9 are crosslinker bead with fewer than four bonds with chain ends and exactly four bonds with chain ends, respectively.
Finally, there are 50 separate filler clusters with 7 icosahedral clusters each, resulting in 7 center beads (type 4), 84 inner-shell beads (type 5), 294 middle-shell beads (type 6), and 644 outer-shell beads (type 7).
Unlike for the polymer and crosslinker molecules, it is important to distinguish between distinct clusters for our desired analysis, and thus filler beads are segregated into separate molecules in \textsf{AMDAT}.
Another feature of \textsf{AMDAT} implicitly shown above is that one can define all polymer chains as a single molecule - this treatment is appropriate for the highly crosslinked elastomer probed here. One could instead treat each strand within the matrix as a separate molecule, although this would introduce complexities surrounding the distinct atom types within each strand on the basis of crosslinking state. 

We use the self-intermediate scattering function (ISFS) to probe how different parts of the filled elastomer relaxes.
Because filler particles are solid, their ISFS do not fully decay over the times we simulate. Interfacial polymer segments are expected to relax far more slowly than bulk segments due to the presence of dynamical gradients near a rigid particle.\cite{richert_dynamics_2011,simmons_emerging_2016,schweizer_progress_2019,broth_polymers_2021}
Spatially resolved ISFS provides a sensitive probe of interphase dynamics and connectivity.
\textsf{AMDAT} enables this type of local analysis via distance-based binning (\texttt{trajectory\_bin\_list}), proximity selections, and region-specific dynamic properties.

We begin by computing the ISFS for the readily available species/types defined in the trajectory, as in Algorithm~\ref{alg:amdat:pnc_isfs_2}.
In Algorithm~\ref{alg:amdat:pnc_isfs_2}, we select polymer, filler, and center lists, and evaluate the ISFS at a wavevector of 7.07 (in dimensionless inverse LJ distance units) for each list.
For ISFS calculations, the wavevector is determined as follows.
Wavenumber bins of index $m$ are centered at positions of $q_m=\frac{\pi}{L}+\frac{m\pi}{L}=\frac{\pi}{L}(1+m)$.
For distinct two-point correlations, $L$ must be the actual dimension of the box to yield accurate values.
For self-correlations such as ISFS, there is no strict limit on $L$.
In this example algorithm, we set $L$ to 12.4825.
The equation above then yields a wavevecotr of 7.07 when $m$ is 26.
In general, this command allows for simultaneous calculation over a range of wavenumber indexes;
here we have set both the lowest and highest indices to 26 to compute values for only a single wavenumber.
The last option is a flag (0 or 1) that tells \textsf{AMDAT} whether to use the full block in its analysis or not.
This only affects time-frame pairs in separate blocks; a value of zero only uses the first time frames in two blocks as pairs, whereas a unit value compares all possible time-frame pairs for a time delay in cross-block analysis.
\begin{lstlisting}[basicstyle=\ttfamily\small, caption={\textsf{AMDAT} PNC ISFS analysis part 2: Creating polymer, filler, and center lists and evaluating ISFS}, label={alg:amdat:pnc_isfs_2}]
create_list polymer
species polymer species xlinkr

isfs ./data/isfs/polymer.isfs.stats 26 26 xyz 12.4385 0
list polymer

create_list filler
species filler

isfs ./data/isfs/filler.isfs.stats 26 26 xyz 12.4385 0
list filler

create_list center
type_system 7

isfs ./data/isfs/center.isfs.stats 26 26 xyz 12.4385 0
list center
\end{lstlisting}

\begin{figure}[!ht]
    \centering
    \includegraphics[]{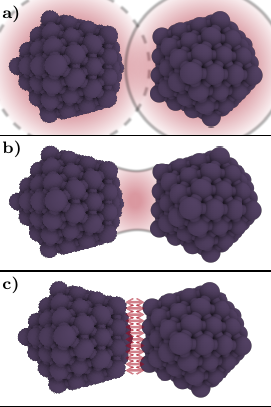}
    \caption{Schematic of the various kinds of dynamically defined local resolution that \textsf{AMDAT} can identify and isolate for analysis: a) Interface, b) Intermolecular regions, and c) Molecular contact.}
    \label{fig:local_res}
\end{figure}

Five regions are identified using \textsf{AMDAT}: a) interfacial polymer, b) second-shell polymer, c) bulk polymer, d) interfiller polymer, and e) contacting filler.
(a), (b), and (c) belong to shells around a filler particle with increasing distance from the surface as per Figure~\ref{fig:local_res}a.
As we will demonstrate, \textsf{AMDAT}'s \texttt{create\_bin\_list} with options \texttt{distance trajectory} is well-suited to identify and isolate such domains.
(d) and (e), shown in Figure~\ref{fig:local_res}b and~\ref{fig:local_res}c, respectively, are identified using different parameterizations of the \texttt{find\_between} command.

\paragraph{Interfacial polymer.}
These are polymer segments that exist in the first shell around filler particles, characterized by a distance of a single bead's diameter separation from the outer shell of the filler particles.
The second and third (bulk) shells are identified similarly (with polymer segments 1-2 and 2-3 bead diameters away from the filler surface, respectively).
The third shell is characterized as the bulk because we expect that dynamic slowing down near the filler surface will extend only two bead diameters away given that these simulations are far above the system's glass transition temperature\cite{lang_interfacial_2013}.
In \textsf{AMDAT}, interface shells are generated via a \texttt{trajectory\_bin\_list} with distance-based binning (3 shells, 1.0 unit thick) anchored on the filler \texttt{outer} list, then converted into named trajectory lists using \texttt{traj\_list\_from\_bin\_list}, as in Algorithm~\ref{alg:amdat:pnc_isfs_3}. 

\begin{lstlisting}[basicstyle=\ttfamily\small, caption={\textsf{AMDAT} PNC ISFS analysis part 3: Create a bin list of polymer segments based on distance from the outer filler shell and then print the ISFS of each shell region.}, label={alg:amdat:pnc_isfs_3}]
create_list outer
type_system 4

# create a bin list called interface_1_3 with 3 shells
# filled with polymer segments based on unit increments
# of distance from outer-shell filler beads
create_bin_list interface_1_3
distance trajectory polymer outer 1.0 3

# create a trajectory list called interface_0-1
# from interface_1_3 with only the first shell
# of polymer segments to target for ISFS analysis
traj_list_from_bin_list interface_0-1 interface_1_3 0 0 0
isfs data/isfs/interface_0-1.isfs.stats 26 26 xyz 12.4385 0
list interface_0-1

traj_list_from_bin_list interface_1-2 interface_1_3 0 0 1
isfs data/isfs/interface_1-2.isfs.stats 26 26 xyz 12.4385 0
list interface_1-2

traj_list_from_bin_list interface_2-3 interface_1_3 0 0 2
isfs data/isfs/interface_2-3.isfs.stats 26 26 xyz 12.4385 0
list interface_2-3
\end{lstlisting}
The syntax for the \texttt{create\_bin\_list distance} command is 
\begin{verbatim}
create_bin_list <name of bin list to create> \n
distance trajectory <list to bin> <list to compute distances from>
<bin thickness> <number of bins>.
\end{verbatim}
For clarity, here the only new line is represented by \textbackslash n. The second and third line would comprise a single line in the input file.

In addition to the \texttt{distance trajectory} option, one can also use the same method to bin into any number of bins in three dimensions based on position (\texttt{all}), position within a rectangular region (\texttt{region}), radial position around a fixed center point (\texttt{distance point}), or based on normal distances and directions from a fixed plane (\texttt{distance plane}).

\begin{figure}[!ht]
    \centering
    \includegraphics[]{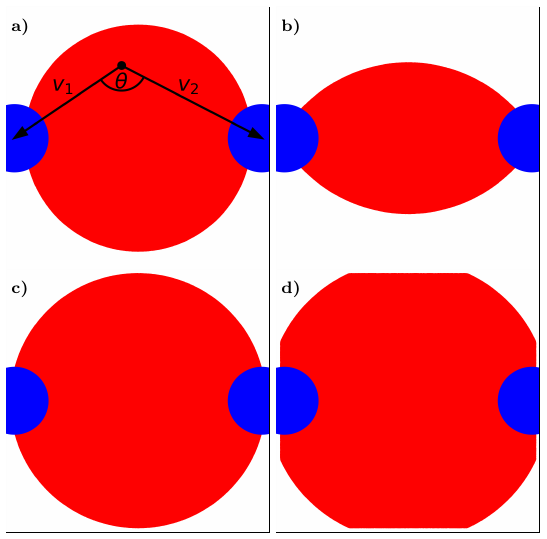}
    \caption{2D illustration of the \texttt{find\_between} command with two particles (blue) separated by a distance of 3. The red-colored region is the selected region by \texttt{find\_between} with ($d_c$, $\cos \theta_c$) options: a) (4.0, 0.0), b) (4.0, -0.5), c) (5.0, 0.0), and d) (5.0, 0.5). $d_c$ and $\cos \theta_c$ are set by the third and fourth arguments of the \texttt{find\_between} command. The labeled point in panel (a) is an example and shows the vectors and angles that \textsf{AMDAT} uses in the selection of particles. Specifically, selected particles will form $\frac{\|\bf{v_1}\|+\|\bf{v_2}\|}{2} < d_c$ and $\cos \theta = \frac{\bf{v_1} \cdot \bf{v_2}}{\|\bf{v_1}\|\|\bf{v_2}\|} < \cos \theta_c$}
    \label{fig:find_between}
\end{figure}

\paragraph{Interfiller polymer.}
In the PNC literature, there is much interest in the specific nature of the interface.
Is it similar to a glassy shell, uniformly distributed around a filler particle, or is it an inter-particle phenomena where polymer segments are impinged by two distinct filler particles?
To study this, we can use the \texttt{find\_between} \textsf{AMDAT} command to isolate polymer segments that live between two distinct filler particles, as illustrated in Figure~\ref{fig:find_between}.
\begin{lstlisting}[mathescape, basicstyle=\ttfamily\small, caption={\textsf{AMDAT} PNC ISFS analysis part 4: Create a trajectory list of polymer segments that reside between two distinct filler particle clusters and evaluate ISFS.}, label={alg:amdat:pnc_isfs_4}]
# create a trajectory list called interfiller with beads from polymer list
# based on two distance vectors with two distinct filler cluster centers
# that sum up to a maximum of 4.25 and have a maximum angle of $\pi$-0.0
find_between interfiller 4.25 0.0 1
list polymer
list center

isfs data/isfs/interfiller.isfs.stats 26 26 xyz 12.4385 0
list interfiller
\end{lstlisting}
The syntax for the \texttt{find\_between} command is \begin{verbatim}
find_between <name of trajectory list to create>
<total magnitude of two distance vectors> <angle of two distance vectors> 
<exclude between the same molecule> \n
<list to fill from>\n 
<list to find between>.
\end{verbatim}

In Algorithm~\ref{alg:amdat:pnc_isfs_4}, the \texttt{trajectory\_list} called \texttt{interfiller} will contain all polymer beads that form vectors with two distinct clusters (\texttt{<exclude between the same molecule>=1}) that sum to less than 4.25$\sigma_{LJ}$ and form an angle whose cosine is less than 0.0.
To accomplish this, \textsf{AMDAT} loops over all polymer beads, computes distance vectors with all (filler) \texttt{center} beads to find the first vector that is lower than 4.25, and then attempts to find a second \texttt{center} bead from distinct filler clusters with a second vector that, together with the first vector, satisfy the criteria above.
The criteria on the angle formed allows exclusion of particles that form distances that sum to less than 4.25 but reside above or below the pair of filler centers.

\paragraph{Contacting filler.}
Another matter of debate in the PNC literature is the role that contacting filler clusters play in load bearing, compression resistance, and dissipation.
Accordingly, we utilize the \texttt{find\_between} command to identify contacting filler beads (also from distinct clusters), as per Algorithm~\ref{alg:amdat:pnc_isfs_5}.
\begin{lstlisting}[mathescape, basicstyle=\ttfamily\small, caption={\textsf{AMDAT} PNC ISFS analysis part 4: Create a trajectory list of filler beads that reside between two distinct contacting filler particle clusters and evaluate ISFS.}, label={alg:amdat:pnc_isfs_5}]
find_between contactfiller 3.75 0.0 1
list outer
list center

isfs data/isfs/contactfiller.isfs.stats 26 26 xyz 12.4385 0
list contactfiller
\end{lstlisting}
By choosing the first target as the outer shell and the second target as the center bead, this command identifies surface beads in filler particles that are in direct contact with another particle.
Because each filler icosahedral particle has a diameter of $\sim 3.5$, we set the distance cut off above to a value of 3.75.
We also limit the choice to surface beads between distinct particles (as we did for the interfiller polymer) to avoid picking up intra-cluster contacts that do not contribute to inter-cluster friction.

As demonstrated by its use for these distinct objectives, the \texttt{find\_between} command is a versatile command that can be used to identify dynamic regions with varying complexity with ease

\begin{figure}[!ht]
    \centering
    \includegraphics[]{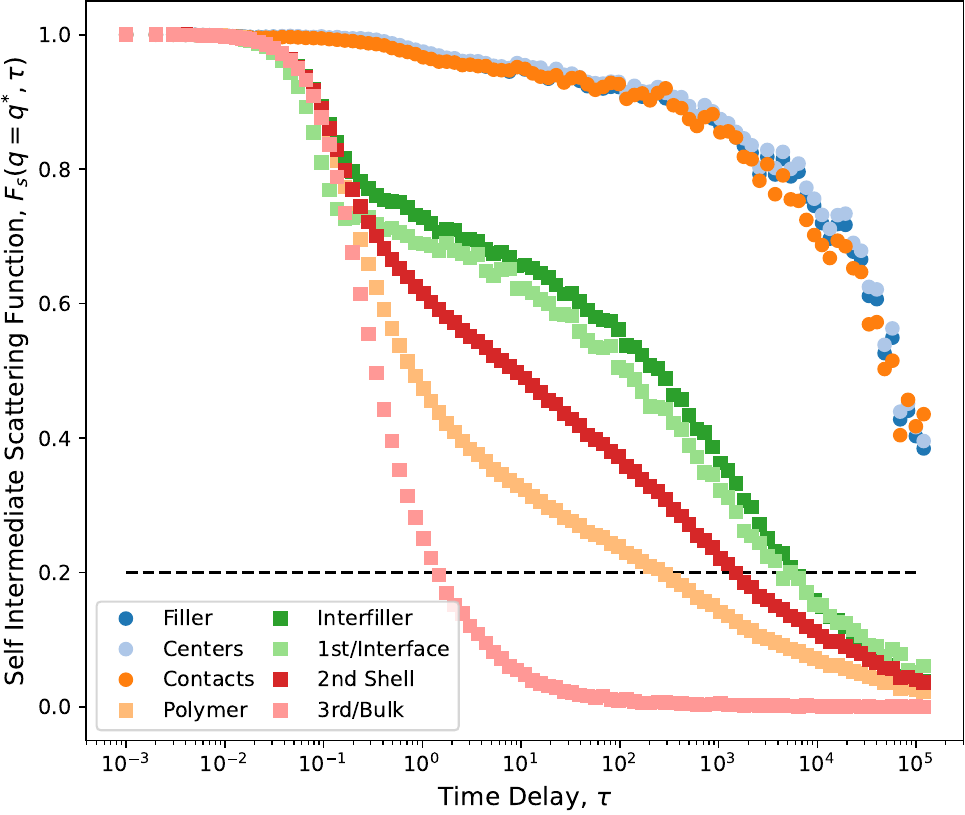}
    \caption{Self-intermediate scattering function $F_s(q^*,t)$ for PNC regions (polymer, filler, centers, interphase shells, interfiller, and contacts).}
    \label{fig:isfs-pnc}
\end{figure}

Put together, Algorithms~\ref{alg:amdat:pnc_isfs_1},~\ref{alg:amdat:pnc_isfs_2},~\ref{alg:amdat:pnc_isfs_3},~\ref{alg:amdat:pnc_isfs_4}, and~\ref{alg:amdat:pnc_isfs_5} form an \textsf{AMDAT} script that selects polymer, filler, and center lists, constructs distance-binned interfacial shells, extracts interfiller and contact subsets, and computes Self Intermediate Scattering Functions (ISFS) for all regions.
The resulting ISFS are graphed in Figure~\ref{fig:isfs-pnc}.
The Filler and Center curves correspond to rigid cluster beads and their geometric centers; as expected for rigid entities, their ISFS does not decay to zero over the simulated window.
The same is true for contacting filler particles and there is no visible difference in the ISFS of these groups.
In contrast, all polymer groups relax within the simulation time, but the rate depends strongly on location: 1st/Interphase ($\texttt{interface\_0-1}$) and Interfiller subsets relax much more slowly than the 3rd/Bulk polymer, consistent with the three-fold stronger polymer--filler attraction that immobilizes chains near the filler surface.
In this system, local proximity to the filler governs relaxation.

\subsubsection{Per-atom property read-in}

The per-atom property read-in method in \textsf{AMDAT}, i.e., the \texttt{custom\_manual} trajectory file type, enables the user to compute averages of properties in different regions during postprocessing, among other capabilities.
Although \textsf{LAMMPS} provides powerful syntax to define different groups of atoms for on-the-fly calculations of spatially averaged properties, its region-based selection is more limited than the geometrically complex calculations reported in the previous section (e.g., interfiller or interface polymer regions in a PNC).
\textsf{AMDAT}'s \texttt{custom\_manual} trajectory file read-in method solves this issue by reading in per-atom properties written by LAMMPS to a trajectory file into a \texttt{value\_list}, which then enables \textsf{AMDAT} to analyze these properties in various ways.
In a recent study, we use this method to show that contacting filler regions exhibit larger normal pressures in a deforming filled elastomer~\cite{Kawak2024}, which mechanistically reinforces the material similar to a pillar holding together a structure by resisting compression.
Here, we demonstrate this capability using the PNC system with a linearly spaced trajectory to obtain per-region counts and average potential energies.

To obtain a trajectory with per-atom properties in \textsf{LAMMPS}, we can employ \textsf{LAMMPS} \texttt{compute} and \texttt{fix ave/atom} commands to prepare the properties and \texttt{dump custom} to print them along with the trajectory.
In Algorithm~\ref{alg:lmp:pnc_custom_manual}, we show the setup that we use in this section to print per-atom potential energies for \textsf{AMDAT} postprocessing.
\begin{lstlisting}[basicstyle=\ttfamily\small, caption={\textsf{LAMMPS} input script to dump per-atom potential energies.}, label={alg:lmp:pnc_custom_manual}]
# compute potential energies per atom
compute pe_atom all pe/atom
# fix ave/atom enables the user to average values over time before dumping;
# Here, the "1 1 200000" configuration samples values every 200000 MD steps
# and does no time averaging
fix pe_atom_avg all ave/atom 1 1 200000 c_pe_atom
# Next, we dump the pe_atom_avg per-atom property using dump custom
dump        dump1 all custom 200000 pe.traj mol type x y z f_pe_atom_avg
dump_modify dump1 append yes sort id
\end{lstlisting}
The setup in Algorithm~\ref{alg:lmp:pnc_custom_manual} will lead to a linearly spaced trajectory, separated by 2$\times 10^5$ MD steps, with columns for the molecule ID, particle type, Cartesian coordinates, and the time-averaged potential energy calculated using \texttt{compute pe/atom} and time-averaged using \texttt{fix ave/atom}.
In the above example, no time averaging is used in the LAMMPS script itself. We employ linear rather than exponential time spacing because only static structural properties, not two-time properties, are studied in this example, and because this approach is sometimes more suitable to a system under deformation.

\begin{figure}[!bp]
    \centering
    \includegraphics[]{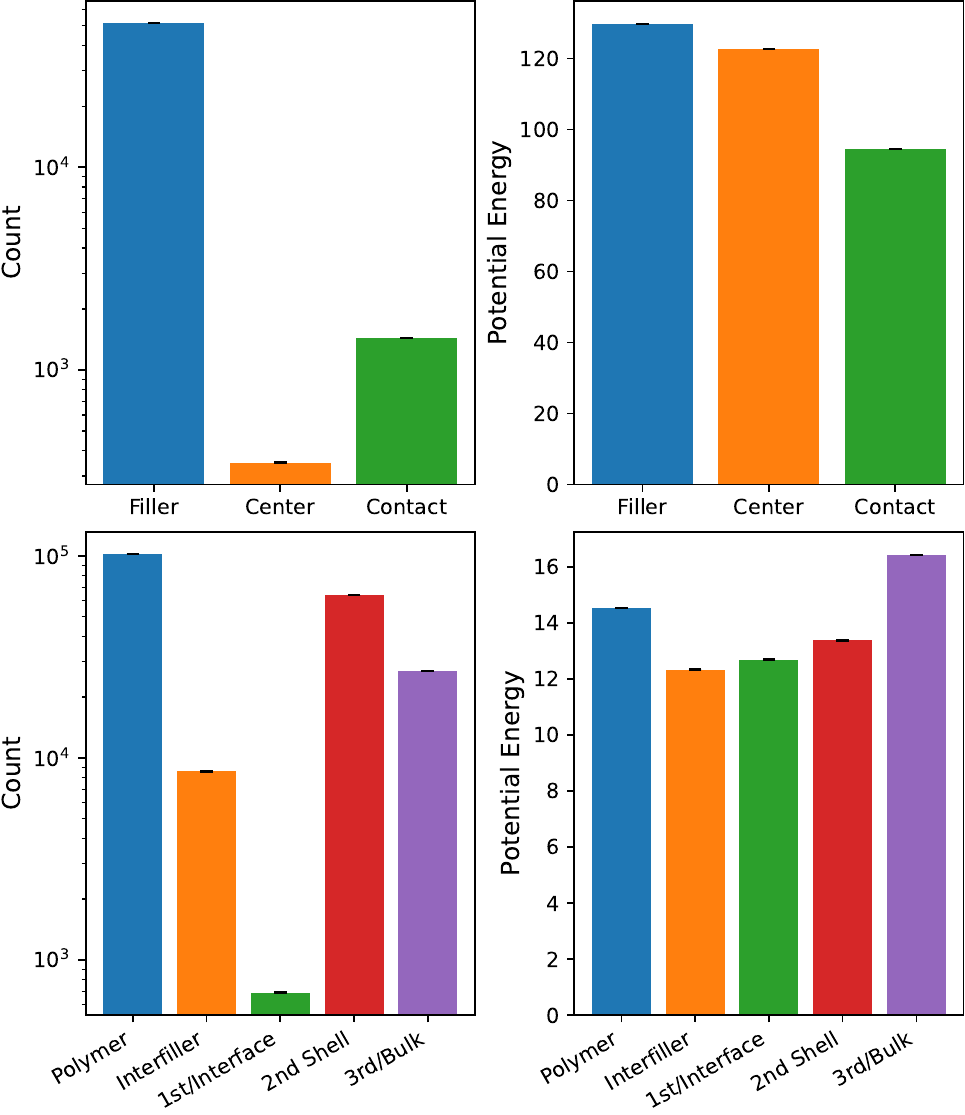}
    \caption{Average counts and potential energies of different subsets/regions within the PNC computed using \textsf{AMDAT}'s \texttt{custom\_manual} read-in method and \texttt{create\_bin\_list}/\texttt{find\_between} methods for identifying contacting filler, interfiller polymer, and polymer shell regions. Each bar has a standard error from the average measurements at different frames.}
    \label{fig:config-pnc}
\end{figure}

Then, with a few changes to the input script in the previous section, \textsf{AMDAT} can read in these per-atom potential energies (PE) and postprocess them to provide the data necessary for Figure~\ref{fig:config-pnc} for the same regions.
We first begin with Algorithm~\ref{alg:amdat:pnc_man_1} by defining the \texttt{custom\_manual} trajectory read-in method.
\begin{lstlisting}[basicstyle=\ttfamily\small, caption={\textsf{AMDAT} PNC \texttt{manual\_config} example part 1}, label={alg:amdat:pnc_man_1}]
system_np
custom_manual
./pe.traj manual_config
linear 41 2000
polymer 1 xlinkr 1 filler 50
1 2 3 4 5 6 7 8 9
90000 500 0 0 0 0 0 9500 0
0 0 326 0 0 0 0 0 2174
0 0 0 644 294 84 7 0 0
\end{lstlisting}
In Algorithm~\ref{alg:amdat:pnc_man_1}, we provide \textsf{AMDAT} with the trajectory and add another file names that maps the columns to \texttt{value\_list} names.
For this example, the \texttt{manual\_config} file is shown in Algorithm~\ref{alg:manual_config}.
\begin{lstlisting}[basicstyle=\ttfamily\small, caption={\texttt{manual\_config} file for mapping trajectory columns to \textsf{AMDAT} \texttt{value\_list} objects}, label={alg:manual_config}]
unwrapped x y z
f_pe_atom_avg pe
\end{lstlisting}
Algorithm~\ref{alg:manual_config} instructs \textsf{AMDAT} to expect unwrapped position coordinates and one additional column called \texttt{f\_pe\_atom\_avg} and to map it to a \texttt{value\_list} called \texttt{pe}.

Then, the remainder of the \textsf{AMDAT} script can refer to \texttt{pe} as a \texttt{value\_list}, providing access to \textsf{AMDAT}'s rich syntax for postprocessing.
For example, to obtain the metrics used in Figure~\ref{fig:config-pnc}, we can use the \texttt{value\_statistics\_pertime} command to obtain a time series of counts, mean, and standard deviation, as shown by Algorithm~\ref{alg:amdat:pnc_man_2}.
Because the analysis here is performed on a quiescent NPT simulation, we further postprocess these space-separated values to averages for each group and plot them in Figure~\ref{fig:config-pnc}.
\begin{lstlisting}[basicstyle=\ttfamily\small, caption={\textsf{AMDAT} PNC \texttt{manual\_config} example part 2: Printing time series of per-atom properties.}, label={alg:amdat:pnc_man_2}]
# get time series of count, PE mean, and PE standard deviation for filler beads
create_list filler
species filler

value_statistics_pertime ./stats/filler.pe.stats pe 3 filler

# get time series of count, PE mean, and PE standard deviation for the
# first shell of polymer around filler particles (interface)
create_bin_list interface_1_3
distance trajectory polymer outer 1.0 3

traj_list_from_bin_list interface_0-1 interface_1_3 0 0 0
value_statistics_pertime ./stats/interface/interface_0-1_center.pe.stats pe 3 interface_0-1

# get time series of count, PE mean, and PE standard deviation for the
# contacting filler region
find_between contactfiller 3.75 0.0 1
list outer
list center

value_statistics_pertime ./stats/contact/contactfiller.pe.stats pe 3 contactfiller
\end{lstlisting}
The syntax for the \texttt{value\_statistics\_pertime} command is \texttt{value\_statistics\_pertime <name of output file> <value list to compute statistics for> <number of moments of distribution to output> <trajectory list to target>}.
This read-in method allows the user to combine \textsf{AMDAT}'s powerful local resolution with property reductions to further characterize the trajectories' regions and extract essential physics.

\begin{figure}[!htbp]
    \centering
    \includegraphics[]{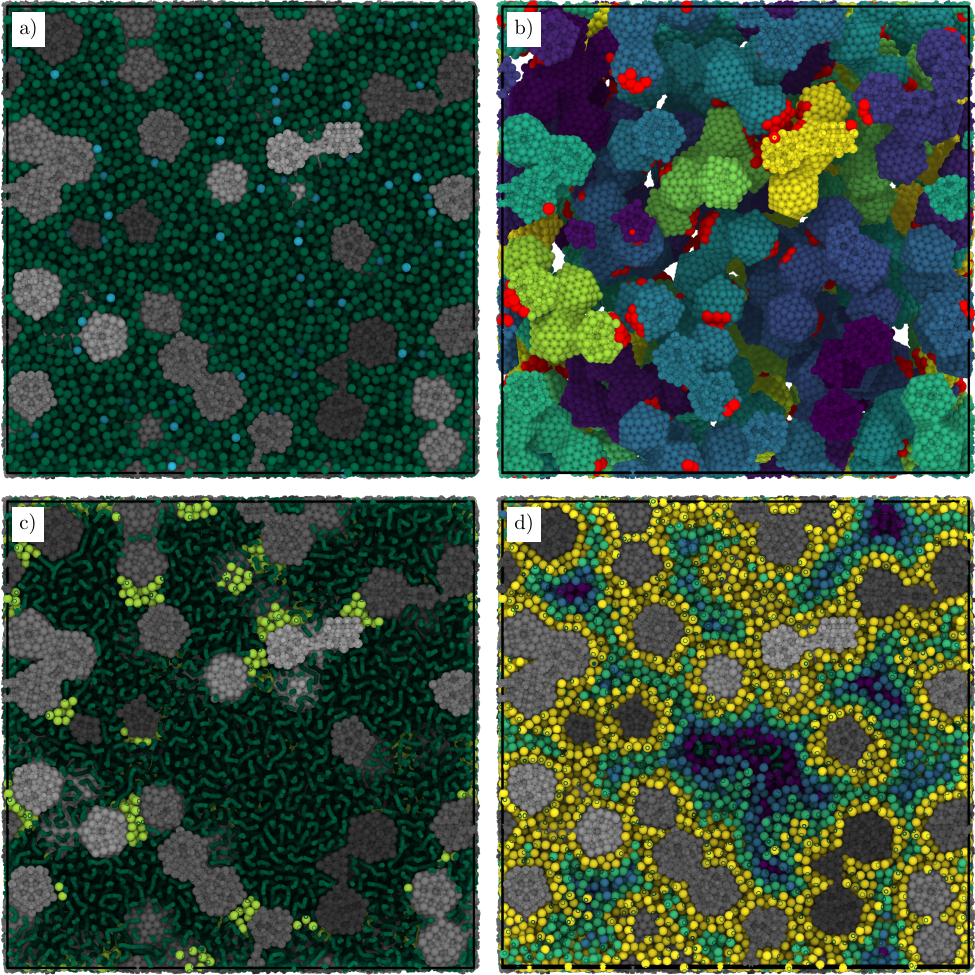}
    \caption{Configurations of PNC system rendered using \textsf{OVITO}~\cite{OVITO}. a) Filler particles illustrated using different colors of gray, polymer beads using green, and crosslinker beads in sky blue. b) Only filler beads are shown with contacting beads (determined using \texttt{find\_between}) shown in red. c) Interfiller polymer, determined using \textsf{AMDAT} \texttt{create\_bin\_list} \texttt{distance} method, are highlighted in apple green. d) Shells around filler particles (\texttt{find\_between}) are shown in different colors.}
    \label{fig:viz-pnc}
\end{figure}

Finally, one can utilize \textsf{AMDAT}'s \texttt{write\_list\_trajectory} command to print the regions to an \texttt{xyz} file for further processing or visualization purposes.
For example, to extract a trajectory of only contacting filler beads at all times, we can utilize the command \texttt{write\_list\_trajectory contactfiller ./xyz/contactfiller.xyz}.
The same command was used to extract trajectories for interfiller beads (also using \texttt{find\_between}) and each of the polymer shell bins (from \texttt{create\_bin\_list}) and we visualize these trajectories using \textsf{OVITO} in Figure~\ref{fig:viz-pnc}.

To accomplish the four visualizations of the same snapshot in the panels of Figure~\ref{fig:viz-pnc}, we loaded the entire trajectory first (\texttt{pe.traj}) and then added (to the \textsf{OVITO} pipeline) each of the \texttt{xyz} trajectory files created using \texttt{write\_list\_trajectory} and distinguished them using color.
Figure~\ref{fig:viz-pnc}a shows the three components in the filled elastomer with polymer in green, crosslinker in cyan, and distinct filler clusters colored in different shades of gray.
Figure~\ref{fig:viz-pnc}b only shows filler beads with contacting filler particles labeled in red.
Figure~\ref{fig:viz-pnc}c shows interfiller polymer regions that reside between two distinct filler clusters.
Finally, Figure~\ref{fig:viz-pnc}d returns to the shades-of-gray coloring scheme for filler clusters but colors polymer beads according to the shell that they reside in; yellow is used for the first two shells with unit thickness and the next three shells are colored using different colors.

Collectively, these examples illustrate \textsf{AMDAT}'s rich ability to enable spatially resolved analysis and visualization.


\section{Multithreading}


\textsf{AMDAT} supports OpenMP CPU multithreading to accelerate post-simulation analysis, achieving significant performance increases. Multithreading is typically implemented simply by parallelizing distinct times or time gaps within static or dynamic calculations, respectively. Figures~\ref{fig:perf-bondacf} and~\ref{fig:perf_isfs} illustrate how multithreading enhances performance of two analyses -- reorientational autocorrelation function and self-intermediate scattering function. Figures~\ref{fig:perf-bondacf}a and~\ref{fig:perf_isfs}a compare the time it takes for \textsf{AMDAT} to run each analysis with increasing number of cores. For each data point, we take the ratio of serial \textsf{AMDAT} versus its multithreaded version to measure speedup. Figures~\ref{fig:perf-bondacf}b and~\ref{fig:perf_isfs}b display the scalability of \textsf{AMDAT} as additional computational resources are allocated. As can be seen from the line graph, \textsf{AMDAT} performs substantially faster as more cores are used, but the performance varies slightly depending on the problem size, as smaller time gaps typically make OpenMP overhead a larger fraction of the total execution time. Although larger workloads generally reduce the relative impact of OpenMP overhead, the measured performance does not always show monotonicity, as illustrated in Figure~\ref{fig:perf-bondacf}b. Each data point is calculated by averaging over multiple runs performed at different instances on a shared HPC cluster. Consequently, the results are subject to fluctuations in system load and resource contention, which introduce variability into the measured performance. Up-to-date performance scaling with multithreading for additional analysis methods can be found on the \textsf{AMDAT} GitHub repository. 

\makeatletter
\newsavebox{\plotpanelbox}
\newcommand{\panelplot}[2]{%
    \sbox{\plotpanelbox}{%
        \includegraphics[
            width=\linewidth,
            height=0.285\textheight,
            keepaspectratio
        ]{#2}%
    }%
    \begingroup
    \setlength{\unitlength}{1pt}%
    \dimen0=0.50\wd\plotpanelbox%
    \dimen2=0.84\ht\plotpanelbox%
    \edef\panelwidth{\strip@pt\wd\plotpanelbox}%
    \edef\panelheight{\strip@pt\ht\plotpanelbox}%
    \edef\panelx{\strip@pt\dimen0}%
    \edef\panely{\strip@pt\dimen2}%
    \begin{picture}(\panelwidth,\panelheight)
        \put(0,0){\usebox{\plotpanelbox}}%
        \put(\panelx,\panely){%
            \makebox(0,0)[c]{\large\bfseries #1)}%
        }%
    \end{picture}%
    \endgroup
}
\makeatother

\begin{figure}[!htbp]
    \centering


    \begin{subfigure}[t]{0.49\textwidth}
        \centering
        \panelplot{a}{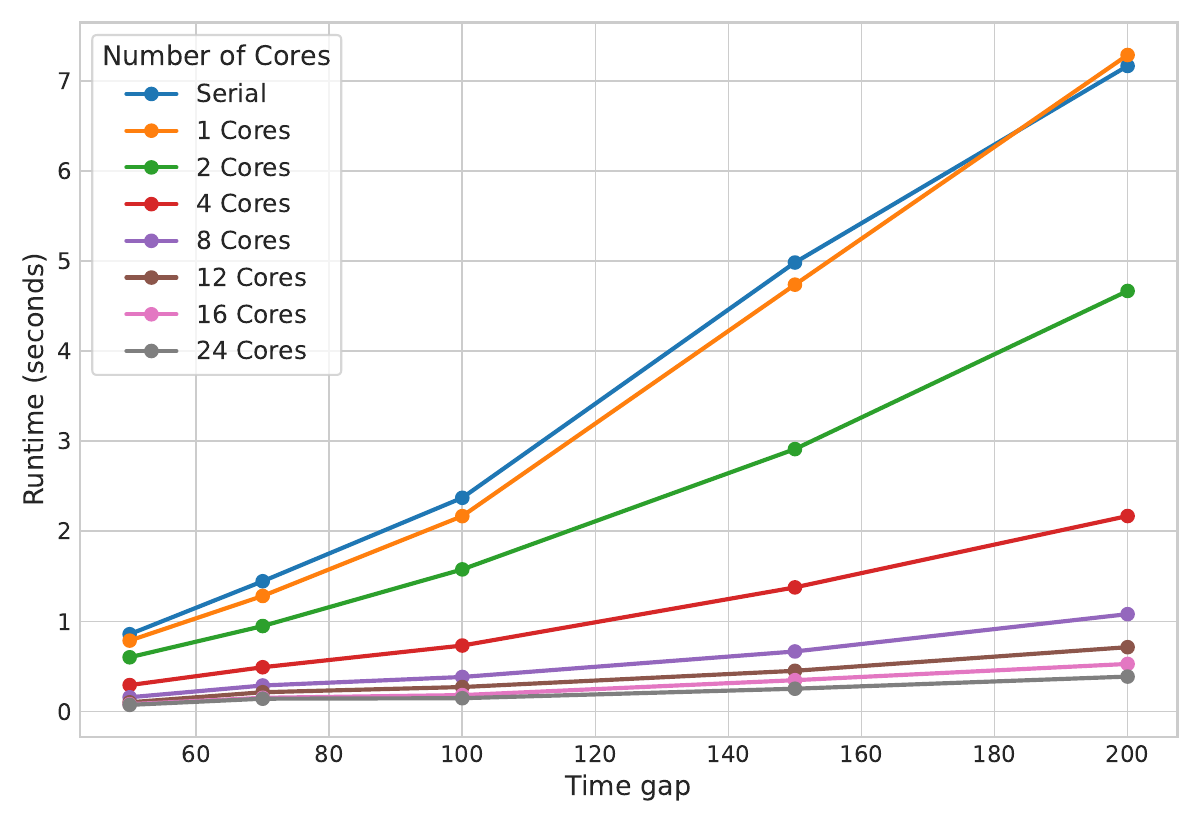}
        \label{fig:baf-runtime}
    \end{subfigure}
    \hfill
    \begin{subfigure}[t]{0.49\textwidth}
        \centering
        \panelplot{b}{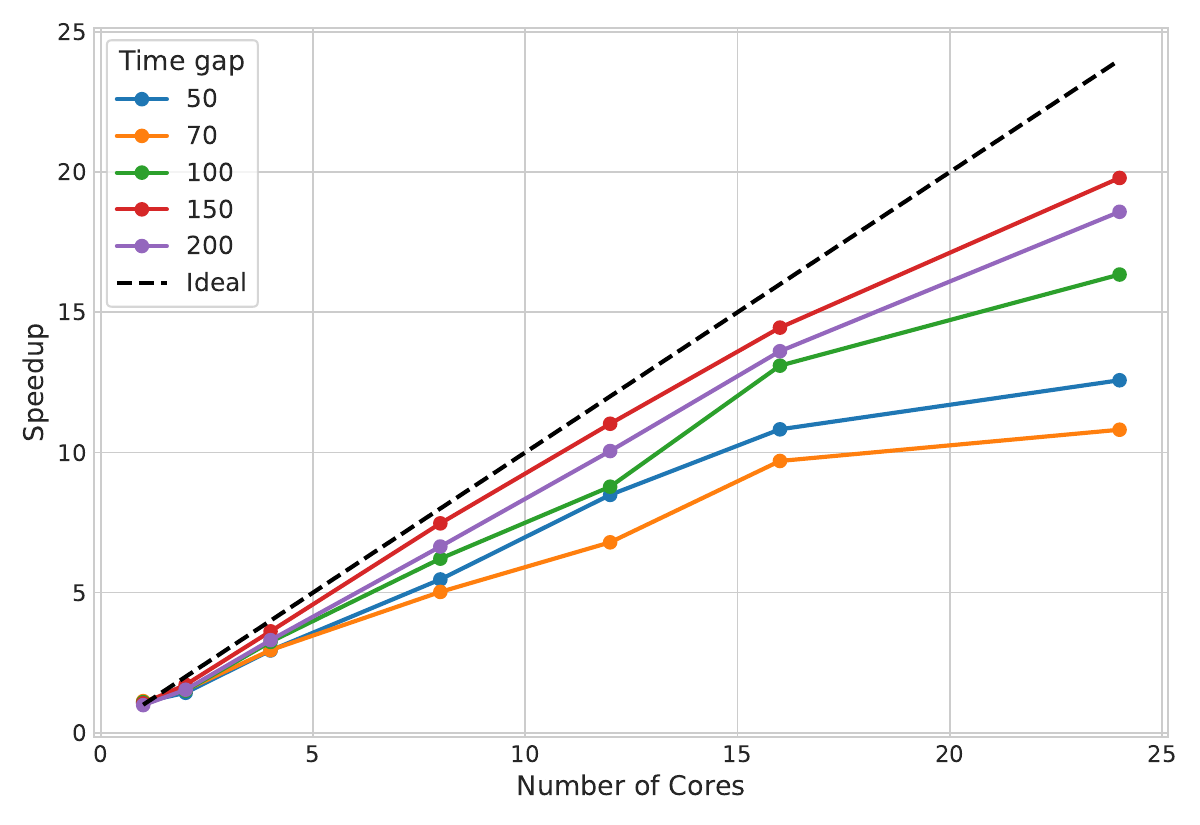}
        \label{fig:baf-speedup}
    \end{subfigure}

    \caption{Performance benchmarks for reorientational autocorrelation function. a) Wall clock time in seconds versus time gaps for serial \textsf{AMDAT} and multithreaded \textsf{AMDAT} runs with different number of cores. b) Speedup of \textsf{AMDAT} versus the number of cores compared to ideal scaling (Speedup = number of processors) for different time gaps.}
    \label{fig:perf-bondacf}
\end{figure}

\begin{figure}[!htbp]
    \centering


    \begin{subfigure}[t]{0.49\textwidth}
        \centering
        \panelplot{a}{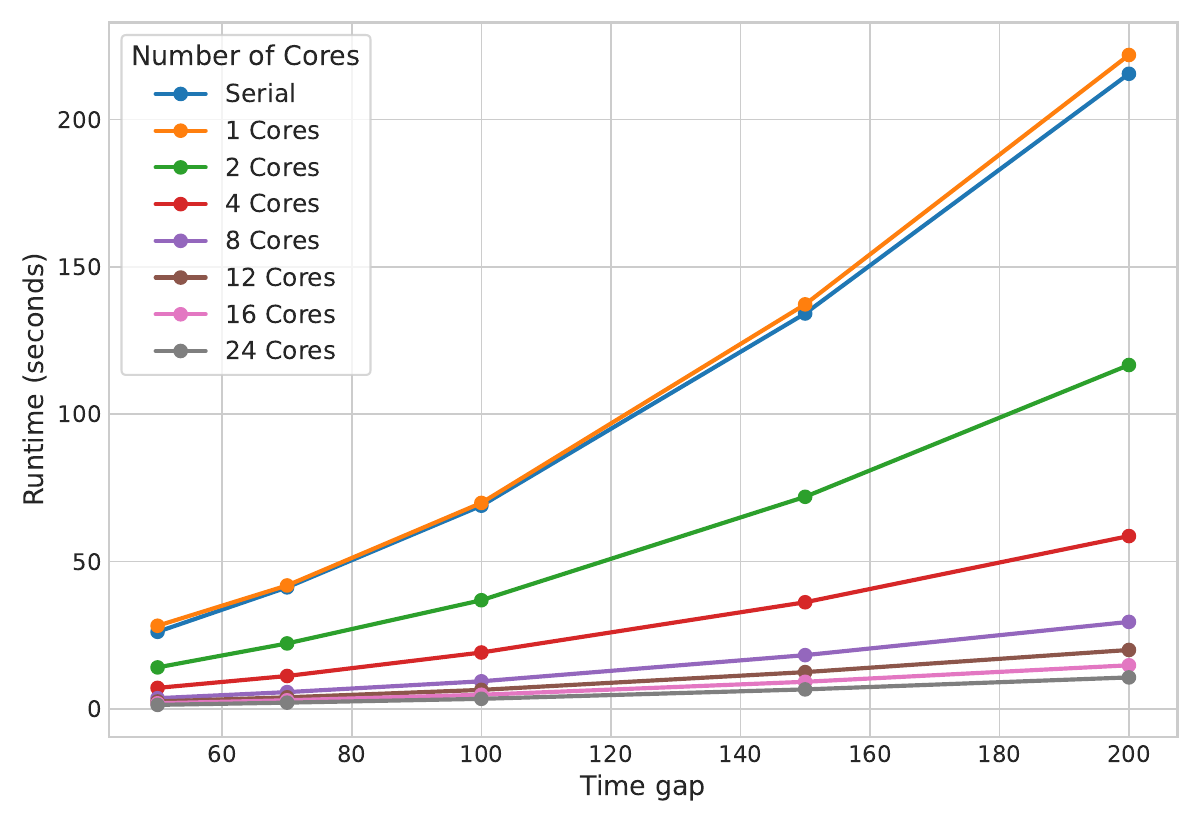}
        \label{fig:isfs-runtime}
    \end{subfigure}
    \hfill
    \begin{subfigure}[t]{0.49\textwidth}
        \centering
        \panelplot{b}{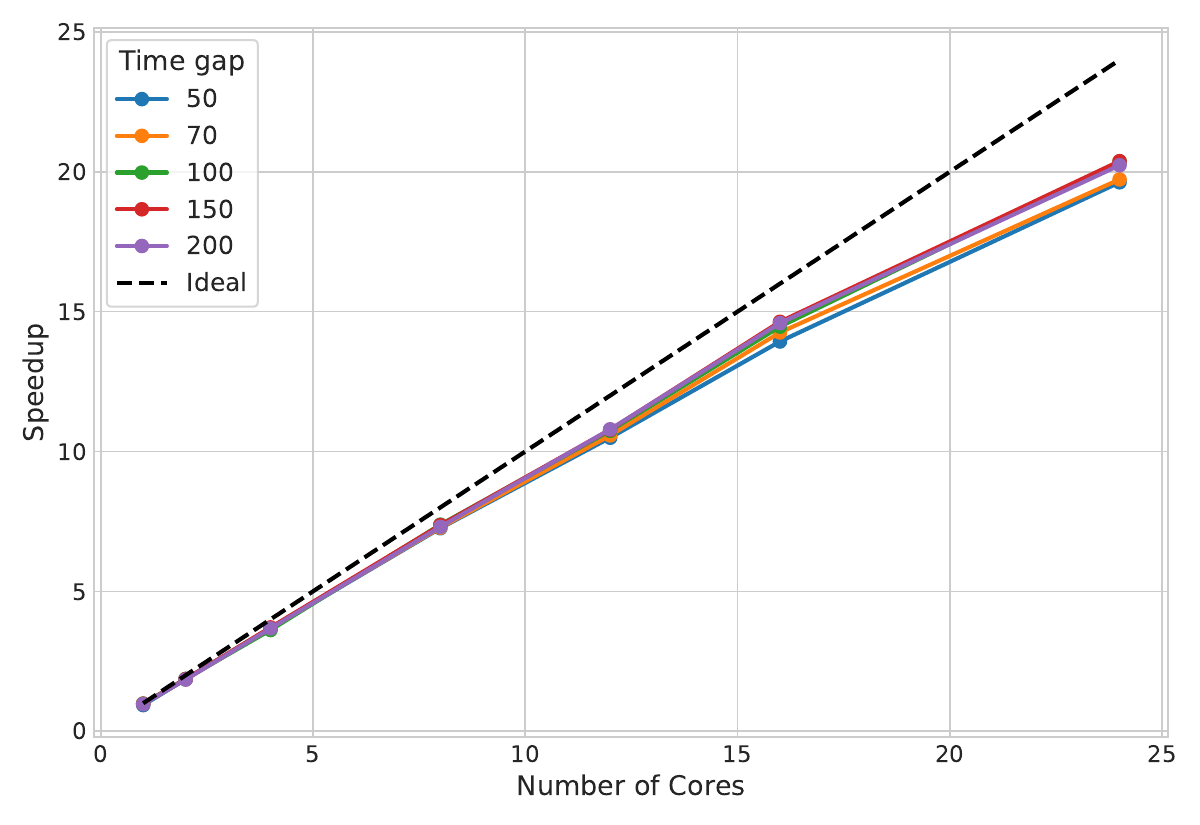}
        \label{fig:isfs-speedup}
    \end{subfigure}

    \caption{Performance benchmarks for self-intermediate scattering function (ISFS). a) Wall clock time in seconds versus time gaps for serial \textsf{AMDAT} and multithreaded \textsf{AMDAT} runs with different number of cores. b) Speedup of \textsf{AMDAT} versus the number of cores compared to ideal scaling (Speedup = number of processors) for different time gaps.}
    \label{fig:perf_isfs}
\end{figure}


\section{Conclusion}


\textsf{AMDAT}, Amorphous Molecular Dynamics Analysis Toolkit, provides a solution for analysis of trajectories of molecular dynamics simulations for polymers, glass-forming, and other soft matter systems, with a focus on amorphous systems and systems where dynamics over many decades of time are of interest.
It incorporates versatile particle-selection capabilities and numerous static and dynamic analysis tools with capabilities in exponential time-spacing and in-memory trajectory analysis to allow for efficient and versatile trajectory analysis.
\textsf{AMDAT} is written in \texttt{C++} in a highly object-oriented manner to facilitate straightforward extensibility of its codebase with additional analysis tools.
It is designed to combine effectively with widely-used simulation software such as \textsf{LAMMPS}, and with popular visualization software such as \textsf{VMD} or \textsf{OVITO}.
For research groups with a focus on molecular dynamics simulation of amorphous systems, \textsf{AMDAT} may thus provide a valuable alternative to in-house analysis tool development or to in-simulation-package analysis capabilities.


\section{Software Availability}


\textsf{AMDAT} is distributed with a GNU GPLv3.0 license as an open-source package.
Installation details, documentation, and examples can be found within the GitHub repository\cite{Simmons_Amorphous_2025} \url{https://github.com/dssimmons-codes/AMDAT}.


\section*{Acknowledgements}


Development of various elements of \textsf{AMDAT} has been supported by multiple grants, including support from the W.~M.~Keck Foundation (early algorithmic improvements), support from National Science Foundation grants DMR - 131043 (early implementation of spatially resolved analysis),  DMR - 1849594 (early architectural improvements), CBET - 1854308 (additions to spatially resolved analysis), and DMR – 2312324 (additions to spatially resolved multibody analysis), support from AFOSR FA9550-22-1-042 (algorithmic improvements and improvements to data read-in), and support from Department of Energy Basic Energy Sciences Grant DE-SC0022329 (implementation of custom read in and other associated capabilities).


\printbibliography

\end{document}